\batchmode
\makeatletter
\def\input@path{{C:/Users/yxk/Desktop/JSAC/my_article_lyx/format_11_28/}}
\makeatother
\documentclass[twocolumn,journal]{IEEEtran}
\pdfoutput=1
\usepackage[T1]{fontenc}
\usepackage[latin9]{inputenc}
\usepackage{float}
\usepackage{booktabs}
\usepackage{calc}
\usepackage{textcomp}
\usepackage{bm}
\usepackage{amsmath}
\usepackage{amsthm}
\usepackage{amssymb}
\usepackage{graphicx}
\usepackage[pdftex,unicode=true,
 bookmarks=true,bookmarksnumbered=true,bookmarksopen=true,bookmarksopenlevel=1,
 breaklinks=false,pdfborder={0 0 0},pdfborderstyle={},backref=false,colorlinks=false]
 {hyperref}
\hypersetup{pdftitle={Your Title},
 pdfauthor={Your Name},
 pdfpagelayout=OneColumn, pdfnewwindow=true, pdfstartview=XYZ, plainpages=false}

\makeatletter

\providecommand{\tabularnewline}{\\}
\floatstyle{ruled}
\newfloat{algorithm}{tbp}{loa}
\providecommand{\algorithmname}{Algorithm}
\floatname{algorithm}{\protect\algorithmname}

\let\oldforeign@language\foreign@language
\DeclareRobustCommand{\foreign@language}[1]{%
  \lowercase{\oldforeign@language{#1}}}
\theoremstyle{plain}
\newtheorem{thm}{\protect\theoremname}
\theoremstyle{remark}
\newtheorem{rem}[thm]{\protect\remarkname}

\usepackage[caption=false,font=normalsize,labelfont=sf,textfont=sf]{subfig}
\usepackage{cite}
\usepackage{bm}
\usepackage{algorithmic}
\usepackage{algorithm}

\usepackage{caption}

\usepackage{graphicx}
\usepackage{epstopdf}
\usepackage[font=footnotesize,labelsep=period]{caption}

\IEEEoverridecommandlockouts

\renewcommand{\algorithmicrequire}{\textbf{Input:}} % Use Input in the format of Algorithm
\@ifundefined{showcaptionsetup}{}{%
 \PassOptionsToPackage{caption=false}{subfig}}
\usepackage{subfig}
\makeatother

\providecommand{\remarkname}{Remark}
\providecommand{\theoremname}{Theorem}

\begin{document}
\title{Robust TOA-based Localization with Inaccurate Anchors for MANET}
\author{Xinkai~Yu,~Yang~Zheng,~\IEEEmembership{Member,~IEEE,}~Min~Sheng,~\IEEEmembership{Senior Member,~IEEE,}~Yan
Shi,~\IEEEmembership{Member,~IEEE,} and~Jiandong~Li,~\IEEEmembership{Fellow,~IEEE}\thanks{Xinkai Yu, Yang Zheng, Min Sheng, Yan Shi and Jiandong Li are with
the State Key Laboratory of Integrated Service Networks, Xidian University,
Xi\textquoteright an 710071, China (e-mail: \protect\href{mailto:xinkaiyu@stu.xidian.edu.cn}{xinkaiyu@stu.xidian.edu.cn};
\protect\href{mailto:yangzheng@xidian.edu.cn}{yangzheng@xidian.edu.cn};
\protect\href{mailto:mshengxd@gmail.com}{mshengxd@gmail.com}; \protect\href{mailto:yshi@xidian.edu.cn}{yshi@xidian.edu.cn};
\protect\href{mailto:jdli@xidian.edu.cn}{jdli@xidian.edu.cn}).}}
\markboth{Journal of XXX}{Your Name \MakeLowercase{\emph{et al.}}: Robust TOA-based Localization
with Inaccurate Anchors for MANET}
\maketitle
\begin{abstract}
Accurate node localization is vital for mobile ad hoc networks (MANETs).
Current methods like Time of Arrival (TOA) can estimate node positions
using imprecise baseplates and achieve the Cram�r-Rao lower bound
(CRLB) accuracy. In multi-hop MANETs, some nodes lack direct links
to base anchors, depending on neighbor nodes as dynamic anchors for
chain localization. However, the dynamic nature of MANETs challenges
TOA's robustness due to the availability and accuracy of base anchors,
coupled with ranging errors. To address the issue of cascading positioning
error divergence, we first derive the CRLB for any primary node in
MANETs as a metric to tackle localization error in cascading scenarios.
Second, we propose an advanced two-step TOA method based on CRLB which
is able to approximate target node's CRLB with only local neighbor
information. Finally, simulation results confirm the robustness of
our algorithm, achieving CRLB-level accuracy for small ranging errors
and maintaining precision for larger errors compared to existing TOA
methods.
\end{abstract}


\IEEEpeerreviewmaketitle{}

\section{Introduction}

\IEEEPARstart{G}{lobal} Positioning System (GPS) is a high-precision
positioning system that relies on satellites in space to achieve comprehensive
global coverage. However, to balance the trade-off of accuracy and
coverage, GPS positioning is not highly accurate in complex environments,
such as tunnel detection scenarios or mountain rescue operations\cite{TOAtunnel,bang3}.
Furthermore, the uncertainties in GPS localization signals are compounded
by the distance between satellites and targets. These uncertainties
arise from various unfavorable signal phenomena\cite{bang5}, including
multipath effects\cite{GPSmultipath}, Non Line-of-Sight (NLOS) connections
between devices\cite{GPSNLOS}, fluctuations in the propagation speed
of radio electromagnetic waves\cite{GPSradiospeed}, and imperfect
synchronization among the system clocks of satellites and targets\cite{GPSsynchronization}.

In GPS-denied or GPS-degraded environments, a mobile ad hoc network
(MANET) that constitutes a small number of base anchors and other
mobile nodes can effectively adapt to the specific attributes of these
restricted areas\cite{LPSuavmanet,localization6G}. Therefore, the
utilization of base anchors to determine the positions of other unlocalized
blind nodes has become a focus in the local position system in MANETs.

Wireless localization techniques can be effectively utilized in MANETs
without relying on external network infrastructure. These algorithms
are classified into two categories based on their approach to measuring
absolute point-to-point distances: range-free and range-based methods\cite{bang1}.

Range-free methods leverage network topology and relative node relationships
through multi-hop transmission. Examples include Distance Vector-Hop
(DV-Hop)\cite{niculescu2003dv}, multidimensional scaling map (MDS-MAP)\cite{MDS},
and others. While these methods reduce hardware demands on nodes and
minimize network overhead, their positioning accuracy is comparatively
lower than that of range-based methods\cite{RangeBased}. In contrast,
range-based methods rely on measurements of physical properties obtained
from neighboring nodes. Techniques such as Time of Arrival (TOA)\cite{TOA},
Time Difference of Arrival (TDOA)\cite{TDOA}, Angle of Arrival (AOA)\cite{AOA},
Received Signal Strength Indication (RSSI)\cite{RSSI}, and and their
combinations are employed. Among these, TOA and TDOA, which are time-based
measurements, are widely utilized due to their optimal balance between
accuracy, stability, ease of hardware implementation, and cost efficiency\cite{LPS}.

TOA methods involve measuring the total flight time of a positioning
signal from an emitter to a receiver, demanding complete synchronization
among all system devices \cite{TOAsynchronization}. Conversely, TDOA
methods measure the relative time difference in signal reception between
two distinct sensors, eliminating the need to synchronize the target
clock with the architecture's sensor clocks\cite{TDOAsynchronization}.
TOA systems offer enhanced adaptability and accuracy in dynamic environments
as they operate independently of baseline information. Additionally,
TDOA methods stem from TOA methods, as the time difference can be
directly derived from the signal's flight time\cite{TDOAandTOA}.
Consequently, we focus our research on the TOA architecture for these
compelling reasons in this paper.

In the existing research on TOA position systems, various optimization
algorithms have been proposed to tackle the node location problem
while minimizing localization uncertainties, such as the Linear Least
Square (LLS) method\cite{TOALS}, Chan algorithm\cite{TOAchan}, Genetic
Algorithms (GA)\cite{TOAga}, Particle Swarm Optimization (PSO)\cite{TOApso},
and Sparrow Search Algorithm (SSA)\cite{locationSSA}. Many of these
studies utilize the Cram�r-Rao Lower Bounds (CRLB), which serves as
a performance metric, setting a lower limit on the variance of any
unbiased estimator\cite{TOACRLB} to characterize the system uncertainties
in node distributions. However, most studies tend to treat the CRLB
solely as a theoretical performance benchmark to evaluate algorithms,
with few integrating it into anchor selection and positioning optimization\cite{Anchorselect,bang4}.
This is mainly because using the CRLB requires prior knowledge of
the target's location, an often unattainable requirement in practical
scenarios. However, In MANETs, nodes might simultaneously serve as
both target points and anchors, providing convenient conditions for
utilizing CRLB.

This paper explores a multi-hop MANET in which network nodes can establish
communication either through single-hop or multi-hop connections.
When the network node is directly connected to the inaccurate base
anchors, the existing TOA localization algorithm can achieve the CRLB
accuracy. In \cite{SDPzou,SDPYANG,SDPmekon,TDOAsynchronization},
the authors use the semidefinite programming (SDP) method to convert
the nonconvex coordinate estimation problem into convex problem. Moreover,
the authors employ intelligent search algorithms to solve estimate
coordinates in \cite{TOAga,TOApso,TOAssa}, also achieving the CRLB
accuracy. In this paper, we employ the PSO algorithm for such node
types, using it as a basis for subsequent node estimation.

In MANETs, direct localization via base anchors is often infeasible.
This necessitates the identification of \textquotedbl dynamic anchor\textquotedbl{}
neighbor nodes, through which localization is conducted in a cascading
fashion. The position information of these anchors is subject to the
CRLB for preceding-level node localization. In this paper, we propose
an advanced two-step TOA-based localization method using the CRLB.
The first step involves using the iterative Chan and PSO method (iChan-PSO)
to optimize dynamic anchor positions and update CRLB values. In the
second step, we combine node mobility and optimized anchors, utilizing
the PSO method to obtain the final result. In this research, these
two steps facilitate the selection of dependable neighbor nodes as
dynamic anchors where base anchors are unobtainable for direct positioning.

In summary, the major contributions of this paper are listed as follows:
\begin{itemize}
\item Given the limited amount of base anchors and potential positional
inaccuracies, we derived the CRLB for the estimated coordinate error
of any node in a multi-level cascading localization scheme within
MANETs. 
\item Our proposed advanced two-step TOA algorithm effectively leverages
local information, including neighbor node's positions and multi-level
CRLBs, to address the issue of poor localization accuracy caused by
large observed errors.
\item We extend the static algorithm by utilizing the continuity of position
estimation where nodes exhibit highly dynamic movements in MANETs.
The proposed dynamic algorithm aims to enhance the robustness of positioning
accuracy in dynamic scenarios.
\end{itemize}
The remainder of this paper is structured as follows. Section II presents
an overview of the localization network model in multi-hop MANETs.
In Section III, we derive the CRLB for cascading TOA localization
within our network model. Section IV discusses our designed TOA localization
method in MANETs. In Section V, we evaluate the performance of our
proposed algorithm and compare it with other TOA-based wireless localization
methods. Finally, we conclude our paper in Section VI.

Main notations are summarized in Table \ref{tab:Notation-List}.

\begin{table}[tbh]
\caption{NOTATION LIST\label{tab:Notation-List}}

\centering{}%
\begin{tabular}{ll}
\toprule 
lowercase $x$ &
scalar\tabularnewline
bold lowercase $\bm{x}$ &
vector\tabularnewline
bold uppercase $\bm{X}$ &
matrix\tabularnewline
$\hat{x}$, $\hat{\bm{x}}$ &
estimate of a variable\tabularnewline
$\left\Vert \boldsymbol{x}\right\Vert $ &
Euclidean norm of a vector\tabularnewline
$\text{tr}\left(\bm{X}\right)$ &
trace of a matrix\tabularnewline
$[\bm{X}]_{u:\upsilon,m:n}$ &
sub-matrix with the $u^{th}$ to $v^{th}$ rows and the $m^{th}$
to\tabularnewline
 &
$n^{th}$ columns\tabularnewline
$[\boldsymbol{X}]_{u,v}$ &
entry at the $u^{th}$ row and $v^{th}$ column of a matrix \tabularnewline
$\mathbb{E}[\cdot]$ &
expectation operator\tabularnewline
$\text{diag}\left(\text{\ensuremath{\cdot}}\right)$ &
diagonal matrix with the elements inside\tabularnewline
$\mathbf{s}_{i}$, $\tilde{\mathbf{s}}_{i}$ &
true coordinates and observed coordinates of the $i^{th}$\tabularnewline
 &
base anchor or dynamic anchor\tabularnewline
$\mathbf{p}^{k}$ &
the target blind node at level-k\tabularnewline
$r_{i}$ &
observed distance between target node $\mathbf{p}^{k}$ and $\mathbf{s}_{i}$\tabularnewline
$r_{ij}$ &
observed distance between $\mathbf{s}_{i}$ and $\mathbf{s}_{j}$\tabularnewline
$D\left(\text{\ensuremath{\cdot}}\right)$, $d\left(\text{\ensuremath{\cdot}}\right)$ &
dynamic anchor set and its count\tabularnewline
$\bm{L}_{i}$  &
set of level-i nodes\tabularnewline
$\mathbf{n}_{k-1}$  &
total count of nodes from the level k-1 and below\tabularnewline
$p_{q}^{k}$ ,$s_{iq}$  &
the $q^{th}$ element of vectors $\mathbf{p}^{k}$and $\mathbf{s}_{i}$\tabularnewline
$\mathcal{F}$ &
Fisher information matrix\tabularnewline
$\sigma_{ci}^{2}$ &
CRLB for the $i^{th}$ dynamic anchor\tabularnewline
$\bm{\theta}$ &
parameter vector\tabularnewline
$\sigma_{i}^{2}$, $\delta_{i}^{2}\bm{I}_{2}$ &
Gaussian variance of distance and base anchors\tabularnewline
$B(\mathbf{\bm{s}},r)$  &
circular neighborhood centered at $\mathbf{\bm{s}}$ with a radius
$r$\tabularnewline
\bottomrule
\end{tabular}
\end{table}

\section{Network Model}

\begin{figure}[tbh]
\begin{centering}
\includegraphics[width=7cm]{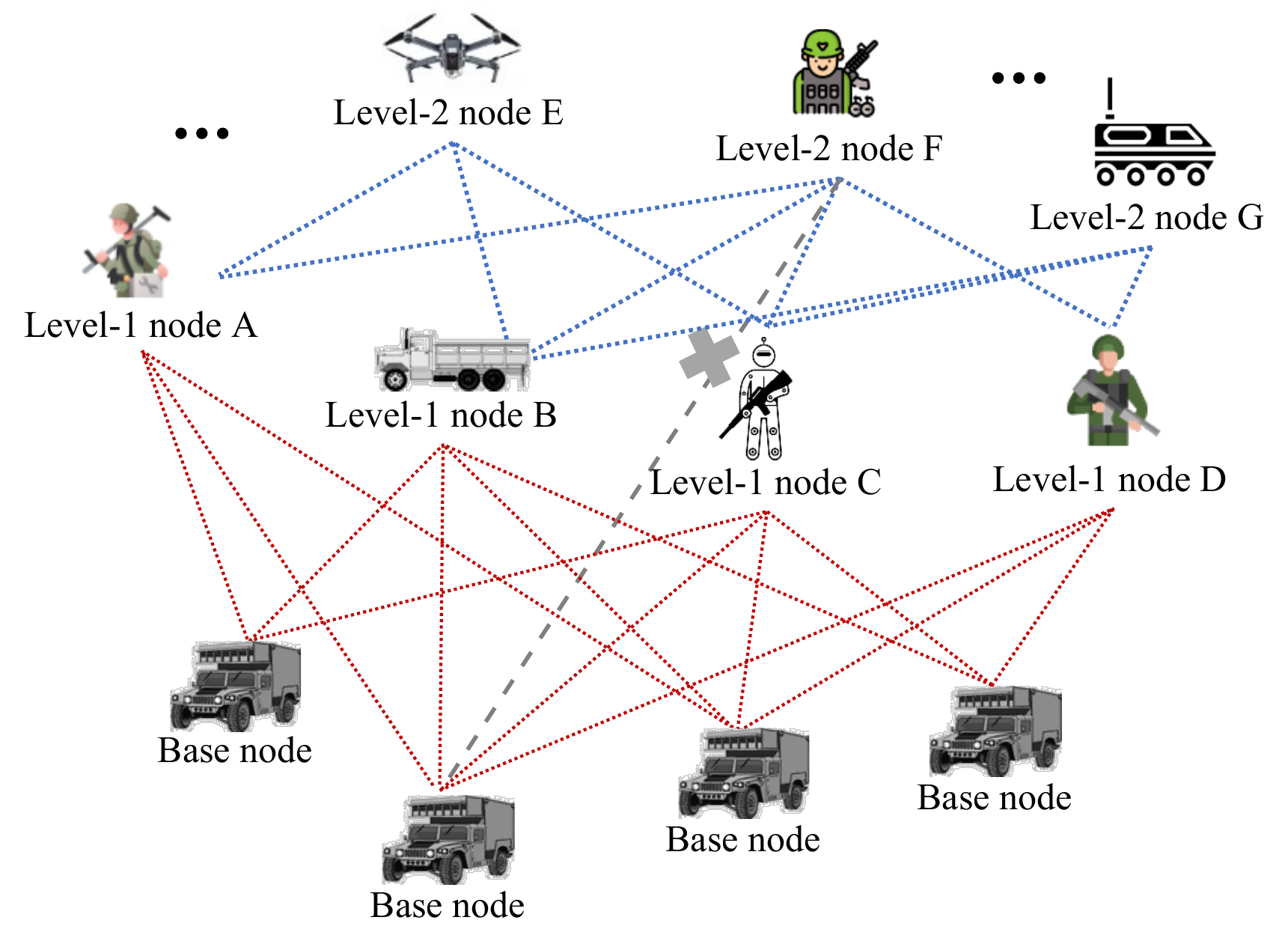}
\par\end{centering}
\caption{Localization system scenario based on TOA in MANET\label{fig:MANET-based-localization-system}}
\end{figure}

In the realm of MANETs, wireless localization systems hold significant
importance across various domains such as emergency response, environmental
monitoring, and bolstering security measures\cite{TOAtunnel,YU20062153,localization6G}.
Figure \ref{fig:MANET-based-localization-system} illustrates a crucial
application in military settings, enabling swift coordination among
personnel and equipment during maneuver warfare. All combatants, communication
vehicles, unmanned aerial vehicles, and other mobile devices function
as network nodes equipped with omnidirectional antennas for signal
transmission and reception. By synchronizing the ad hoc nodes, receiving
units determine transmission times, enabling distance calculations
between nodes based on data signal transmission times.

A hierarchical approach is a common strategy for managing and organizing
network nodes based on functionality, capabilities, resources, and
contributions\cite{dawes2022sdg,bang2}. This paper employs a hierarchical
methodology for node localization, focusing on the transmission radius
between each node in the network, as depicted in Fig. \ref{fig:MANET-based-localization-system}.
Within this method, four imprecise base anchors, possessing prior
positional knowledge, serve as level-0 nodes. Typically, these anchors
require manual deployment in locations like rear command centers or
control consoles for practical implementation. In TOA localization
within an m-dimensional scenario, a minimum of m+1 anchor points is
necessary for accurate localization. Consequently, level-1 nodes need
at least m+1 neighboring base anchors, such as A, B, C, and D, enabling
direct localization from categorized level-0 base anchors.

Nodes beyond the second level face an insufficient number of neighbor
base anchors for direct localization, prompting reliance on lower-level
nodes for support. In this context, level-2 nodes utilize level-1
nodes and partial level-0 nodes from single-hop neighbors, serving
as \textquotedbl dynamic anchors\textquotedbl{} to enhance location
information within our framework. For instance, node F's dynamic anchor
set would comprise \{A, B, C, D\}.

To generalize, nodes at any level can leverage neighbors from lower
levels for localization. Consider a level-k node; it can utilize dynamic
anchors from level k-1 and below, provided at least one level k-1
node acts as a dynamic anchor. To maintain order in localization,
it is not allowed for the target point and its dynamic anchor to be
at the same level. This approach guarantees each node contributes
to the network's position estimation. In practice, node localization
typically proceeds from lower to higher levels, with the weakening
of connections to base anchors as the level increases.

We are dealing with a network including $n_{0}$ level-0 nodes with
prior positional information that serve as base anchors. All network
nodes, including base anchors, are subject to dynamic positional changes,
and the initial position information of base anchors is imprecise.
For simplicity, we consider a two-dimensional scenario, it is easy
to extend for three-dimensional case. Let $\mathbf{s}_{i}^{0}=\left(x_{i},y_{i}\right)^{\text{T}}$
represents the true but unknown coordinates of the $i^{th}$ base
anchor. We assume that $\mathbf{\bm{\beta}}_{i}$ is a zero-mean white
Gaussian vector with known variance $\delta_{i}^{2}\bm{I}_{2}$. The
positions of the base anchors can be represented as
\begin{equation}
\tilde{\mathbf{s}}_{i}^{0}=\mathbf{s}_{i}^{0}+\mathbf{\bm{\beta}}_{i},~i=1,2,\ldots,n_{0}.\label{eq:=0057FA=0051C6=00951A=0070B9=005B58=005728=009AD8=0065AF=008BEF=005DEE}
\end{equation}

Let $\mathbf{p}\bm{=}\left(x,y\right)^{\text{T}}$ be the coordinates
of an unknown target node. Define $d_{i}=\left\Vert \mathbf{p}-\bm{s}_{i}\right\Vert $
as the unknown distance between $i^{th}$ available dynamic anchor
and the target node. Assuming line-of-sight signal propagation, the
asynchronous TOA measurements can be denoted as
\begin{equation}
r_{i}=d_{i}+n_{i},~i=1,2,\ldots,N.\label{eq:=0090BB=005C45=008DDD=0079BB=0052A0=004E0A=009AD8=0065AF}
\end{equation}
where $N$ represents the number of dynamic anchors for unknown target
$\mathbf{p}$. Here, $n_{i}\sim\mathcal{N}\left(0,\sigma_{i}^{2}\right)$
represents the Additive Gaussian White Noise (AWGN) with a variance
$\sigma_{i}^{2}$. In MANETs, network nodes exhibit the capability
to dynamically adjust their behavior, configurations, and connection
patterns in response to changing network conditions, thereby catering
to varying environments and requirements\cite{MANETlocation}. For
any target node, its available dynamic anchors can be any node other
than itself, determined by the network topology and node level. Therefore,
the node localization challenge within MANETs can be segmented into
two main aspects: the localization of level-1 nodes, and the localization
of nodes at level-2 or higher. How to utilize inaccurate base anchors
and dynamic anchors to localize other blind nodes is the focus of
our research. 

\section{Cram�r-Rao Lower Bound for TOA-based Localization in MANETs\label{sec:Cram=0000E9r=002013Rao-Lower-Bound}}

The CRLB, obtained from the inverse of the Fisher Information Matrix
(FIM), represents the minimum achievable error in positioning estimation,
independent of the positioning algorithm employed. Numerous studies
focus on CRLB in TOA localization errors\cite{TOAnewcrlb,TOACRLB}.
However, most derivations only take into account the direct connection
between the target source and base anchors. In MANETs, a large number
of nodes cannot directly connect to base anchors and instead rely
on multi-hop connections for indirect localization. In this section,
we i provide an exposition on the characterization of the CRLB for
nodes at any level under the TOA measurement model and network hierarchy,
which serve as both evaluation criteria and guiding principles for
designing TOA-based localization in MANETs.

Assume that the observed signal $x(t)$ contains information related
to the real parameter $\theta$. For a sample vector, the corresponding
estimator $\hat{\theta}$ can be derived from a single sample vector
$\bm{x}=\left\{ x_{1},x_{2},...,x_{n}\right\} ^{\text{T}}$. Considering
an unbiased estimator, the Cram�r-Rao bound theorem establishes a
necessary and sufficient condition for it to be an optimal estimator.
Treating $\bm{x}$ as a random variable, the conditional probability
density function $f\left(\bm{x}|\theta\right)$ can be constructed
based on the real parameter $\theta$. The Cram�r-Rao bound provides
the lower limit for the mean square error (MSE) of the estimator $\hat{\theta}$,
which can be given by
\begin{equation}
\sigma_{\theta}^{2}=\text{\ensuremath{\mathbb{E}}}\left[(\hat{\theta}-\theta)^{2}\right]\ge\frac{1}{\text{J}(\theta)}\label{eq:fi=00548C=0065B9=005DEE=004E0D=007B49=005F0F-1}
\end{equation}
where $J(\theta)$ represents the FIM, and $\sigma_{\theta}^{2}$
signifies the MSE of estimator $\hat{\theta}$. The elements of the
FIM are calculated through the second derivative of the logarithm
of the likelihood function concerning the parameters, and can be concisely
expressed as
\begin{equation}
\text{J}(\theta)=-\mathbb{E}\left\{ \frac{\partial^{2}}{\partial\theta^{2}}\ln f(\bm{x}|\theta)\right\} 
\end{equation}

In the context of TOA-based positioning in MANETs, each node obtains
range estimates to its neighbor nodes within the communication range.
We consider a MANET includes $n_{0}$ base anchors which server as
level-0 nodes and the number of nodes from level $1$ to the highest
level m is denoted as $n_{1},n_{2},...,n_{m}$, respectively. Meanwhile,
we designate the set of level-i nodes as $\bm{L}_{i}$, with a corresponding
node count of $n_{i}$. Consider the unknown target node at level
k, where k is greater than or equal to $1$. For target node $\mathbf{p}^{k}$,
let $D\left(\mathbf{p}^{k}\right)$ be the dynamic anchor set, representing
a combination of anchors for localization and can be written as
\begin{equation}
D\left(\mathbf{p}^{k}\right)=\left\{ \mathbf{s}_{i}\Bigg|\left\Vert \mathbf{p}^{k}-\mathbf{s}_{i}\right\Vert \leqslant R,\:\mathbf{s}_{i}\in\sum_{i=0}^{\begin{smallmatrix}k-1\end{smallmatrix}}\bm{L}_{i}\}\right\} 
\end{equation}
where $d\left(\mathbf{p}^{k}\right)=\left|D\left(\mathbf{p}^{k}\right)\right|$
denotes the number of dynamic anchors, $R$ is the communication radius.
For simplicity, we set $d\left(\mathbf{p}^{k}\right)=N$ in the subsequent
derivations.

To determine the CRLB under the general TOA measurement model in MANETs,
we assume that the measurement noises in \eqref{eq:=0057FA=0051C6=00951A=0070B9=005B58=005728=009AD8=0065AF=008BEF=005DEE}
and \eqref{eq:=0090BB=005C45=008DDD=0079BB=0052A0=004E0A=009AD8=0065AF}
are both independent and identically distributed (i.i.d.). Gaussian
random variables of base anchor positions and observed distances with
zero mean and variance $\delta^{2}\bm{I}_{2}$ and $\sigma^{2}$.
Define $\bm{\theta}=\left(\mathbf{p}^{k},\bm{L}_{k-1},\bm{L}_{k-2},...,\bm{L}_{0}\right)$
is the estimator in the Maximum Likelihood Estimator (MLE). Under
the i.i.d. assumption, the joint conditional probability density function
of the observed distance is defined as
\begin{equation}
p(\bm{\theta})\triangleq p(r_{ij},r_{i},\tilde{\bm{L}}_{0}|\mathbf{p}^{k},\bm{L}_{0},...,\bm{L}_{k-1})
\end{equation}
where $r_{i}$ denotes the observed distance between the target node
$\mathbf{p}^{k}$ and its $i^{th}$ dynamic anchors. Considering the
observed errors in the formula \eqref{eq:=0057FA=0051C6=00951A=0070B9=005B58=005728=009AD8=0065AF=008BEF=005DEE}
and \eqref{eq:=0090BB=005C45=008DDD=0079BB=0052A0=004E0A=009AD8=0065AF},
$p(\mathbf{\bm{\theta}})$ can be rewritten as
\begin{align}
p(\bm{\theta})= & \prod_{i\,=\,1}^{\begin{smallmatrix}N\end{smallmatrix}}\left[\frac{1}{\sqrt{2\pi\sigma^{2}}}\exp\left(-\frac{\left(r_{i}-\|\mathbf{p}^{k}-\mathbf{s}_{i}\|\right)^{2}}{2\sigma^{2}}\right)\right]\times\nonumber \\
 & \prod_{i\,=\,1}^{\mathbf{n}_{k-1}}\prod_{j\,=\,1}^{\begin{smallmatrix}d\left(\mathbf{s}_{i}\right)\end{smallmatrix}}\left[\frac{1}{\sqrt{2\pi\sigma^{2}}}\exp\left(-\frac{\left(r_{ij}-\|\mathbf{s}_{i}-\mathbf{s}_{j}\|\right)^{2}}{2\sigma^{2}}\right)\right]\times\nonumber \\
 & \prod_{i\,=\,1}^{\mathbf{n}_{0}}\left[\frac{1}{\sqrt{2\pi\delta^{2}}}\exp\left(-\frac{\|\mathbf{s}_{i}^{0}-\tilde{\mathbf{s}}_{i}^{0}\|^{2}}{2\delta^{2}}\right)\right]
\end{align}
where $\mathbf{n}_{k-1}$ represents the total number of network nodes
from the level k-1 and below. Ignoring the constant term, the log-likelihood
function of the location estimation is given by
\begin{align}
\text{L}(\bm{\theta})= & -\frac{1}{2\sigma^{2}}\sum_{i\,=\,1}^{\begin{smallmatrix}N\end{smallmatrix}}\left(r_{i}-\|\mathbf{p}^{k}-\mathbf{s}_{i}\|\right)^{2}\nonumber \\
 & -\frac{1}{2\sigma^{2}}\sum_{i\,=\,1}^{\mathbf{n}_{k-1}}\sum_{j\,=\,1}^{\begin{smallmatrix}d\left(\mathbf{s}_{i}\right)\end{smallmatrix}}\left(r_{ij}-\|\mathbf{s}_{i}-\mathbf{s}_{j}\|\right)^{2}\label{eq:=007EA7=008054=004F3C=007136=0051FD=006570}\\
 & -\frac{1}{2\delta^{2}}\sum_{i\,=\,1}^{\mathbf{n}_{0}}\|\mathbf{s}_{i}^{0}-\tilde{\mathbf{s}}_{i}^{0}\|^{2}.\nonumber 
\end{align}

Therefore, the dimensionality of the FIM is $2\text{\ensuremath{\left(\mathbf{n}_{k-1}+1\right)}}$
in the two-dimensional scenario consistent with estimator $\bm{\theta}$.
From \eqref{eq:=007EA7=008054=004F3C=007136=0051FD=006570}, we can
determine the $\left(m,n\right)$th element in the FIM as
\begin{equation}
\left[\mathcal{F}\right]_{m,n}=\sum_{i\,=\,1}^{N}\frac{(p_{m}^{k}-s_{im})(p_{n}^{k}-s_{in})}{\sigma^{2}\|\mathbf{p}^{k}-\bm{s}_{i}\|^{2}},1\leq m\leq n\leq2
\end{equation}
Similarly, for $1\leq m\leq2<n\leq2\left(\mathbf{n}_{k-1}+1\right)$,
define $n=2n_{1}+n_{2}$, we have
\begin{equation}
\left[\mathcal{F}\right]_{m,n}=\frac{(p_{m}^{k}-s_{n_{1}m})(p_{n_{2}}^{k}-s_{n_{1}n_{2}})}{\sigma^{2}\|\mathbf{p}^{k}-\bm{\mathbf{s}}_{n_{1}}\|^{2}}
\end{equation}
Additionally, for $3\leq m\leq n\leq2\left(\mathbf{n}_{k-1}+1\right)$,
define $m=2m_{1}+m_{2}$ and $n=2n_{1}+n_{2}$, we have
\begin{equation}
\begin{smallmatrix}\left[\mathcal{F}\right]_{m,n}=\end{smallmatrix}\left\{ \begin{aligned} & \frac{\left(s_{m_{1}m_{2}}-s_{n_{1}m_{2}}\right)\left(s_{m_{1}n_{2}}-s_{n_{1}n_{2}}\right)}{\sigma^{2}\|\bm{\mathbf{s}}_{m_{1}}-\bm{\mathbf{s}}_{n_{1}}\|^{2}},m_{1}\neq n_{1}\\
 & \sum_{i=1}^{\begin{smallmatrix}d\left(\bm{\mathbf{s}}_{m1}\right)\end{smallmatrix}}\frac{\begin{smallmatrix}(s_{m_{1}m_{2}}-s_{im_{2}})(s_{m_{1}n_{2}}-s_{in_{2}})\end{smallmatrix}}{\sigma^{2}\|\bm{\mathbf{s}}_{m_{1}}-\bm{\mathbf{s}}_{i}\|^{2}},\begin{smallmatrix}m_{1}=n_{1}\leq\mathbf{n}_{k-1}-\mathbf{n}_{0}\end{smallmatrix}\\
 & \text{\ensuremath{\sum_{i=1}^{\begin{smallmatrix}d\left(\bm{\mathbf{s}}_{m1}\right)\end{smallmatrix}}\frac{\left(s_{m_{1}m_{2}}-s_{im_{2}}\right)^{2}}{\sigma^{2}\|\begin{smallmatrix}\bm{\mathbf{s}}_{m_{1}}-\bm{\mathbf{s}}_{i}\end{smallmatrix}\|^{2}}+\frac{1}{\delta^{2}},\begin{array}{c}
\begin{smallmatrix}\mathbf{n}_{k-1}-\mathbf{n}_{0}\leq m_{1}\leq\mathbf{n}_{k-1}\end{smallmatrix}\\
\begin{smallmatrix},m_{1}=n_{1},m_{2}=n_{2}\end{smallmatrix}
\end{array}}}
\end{aligned}
\right.
\end{equation}
where $m_{2},n_{2}\in\left\{ 1,2\right\} $. The detailed derivation
of the FIM can be found in the Appendix. Hence, the CRLB of the unbiased
estimate $\mathbf{p}^{k}$ is 
\begin{equation}
\sigma_{c}^{2}=\text{tr}\{\left[\mathcal{F}^{-1}\right]_{1:1,2:2}\}=\sum_{i=1}^{2}\left[\mathcal{F}^{-1}\right]_{i,i}\label{eq:TOACRLB}
\end{equation}

\section{Robust TOA-based Localization Algorithm for MANETs}

In this section, we design a TOA-based localization algorithm to make
it suitable for dynamic scenarios in MANETs. For level-1 nodes which
can directly utilize base anchors for localization, we employ existing
algorithms to achieve localization error close to the CRLB level\cite{SDPYANG,SDPzou,SDPmekon,TOApso}.
For level-2 and higher-level nodes, due to the lack of sufficient
directly connected base anchors, we propose an advanced TOA-based
positioning algorithm that utilizes dynamic anchors while considering
the CRLB. Our designed iCHAN-PSO method achieves the correction of
dynamic anchor positions and updates the corresponding CRLB weights.
Additionally, our two-step TOA method integrates node mobility to
address more complex network scenarios.

\subsection{TOA Localization Algorithm for Level-1 Nodes}

For level-1 nodes, their neighbors possess a sufficient number of
base anchors for localization. Thus, the level-1 node position problem
can be transformed into source localization with anchor position uncertainties,
without considering the multi-hop conditions in MANETs. Under the
mutually independent Gaussian noise of base anchor positions and observed
distance, we can write the MLE of one level-1 node as follows
\begin{equation}
\min_{\mathbf{p}^{1},\mathbf{s}_{i}^{0}}\sum_{i\,=\,1}^{d\left(\mathbf{p}^{1}\right)}\frac{(r_{i}-\|\mathbf{p}^{1}-\mathbf{s}_{i}^{0}\|)^{2}}{\sigma^{2}}+\sum_{i\,=\,1}^{d\left(\mathbf{p}^{1}\right)}\frac{\|\tilde{\mathbf{s}}_{i}^{0}-\mathbf{s}_{i}^{0}\|^{2}}{\delta^{2}}\label{eq:level-1=0076EE=006807=0051FD=006570}
\end{equation}
where $\mathbf{p}^{1}$ and $\mathbf{s}_{i}^{0}$ are the optimization
parameters of one level-1 target node and the $i^{th}$ base anchors.
The above problem is nonlinear and nonconvex, and the MLE is hard
to achieve. In this paper, we employ the PSO algorithm to obtain estimates
for these level-1 nodes. Similar to source localization with anchor
position uncertainties, it can be proven that the level-1 node localization
achieves the CRLB accuracy in \cite{TOApso}.

\subsection{TOA Localization for Higher-Level MANET Nodes}

For positioning nodes at level 2 or above, the simplest approach is
to utilize the observed distances between all neighbor nodes in network.
Take the positioning of level-2 nodes as an example. To achieve a
global solution, the MLE of $\mathbf{p}^{2}$ can be constructed as
\begin{equation}
\begin{aligned}\min_{\substack{\mathbf{p}^{2},\mathbf{p}_{i}^{1},\mathbf{s}_{i}^{0}}
}~ & \sum_{i\,=\,1}^{\begin{smallmatrix}d\left(\mathbf{p}^{2}\right)\end{smallmatrix}}\frac{(r_{i}-\|\mathbf{p}^{2}-\mathbf{s}_{i}\|)^{2}}{\sigma^{2}}+\sum_{i\,=\,1}^{\begin{smallmatrix}\mathbf{n}_{0}\end{smallmatrix}}\frac{\|\tilde{\mathbf{s}}_{i}^{0}-\mathbf{s}_{i}^{0}\|^{2}}{\delta^{2}}+\\
 & \sum_{i\,=\,1}^{\begin{smallmatrix}d\left(\mathbf{p}^{1}\right)\end{smallmatrix}}\sum_{j\,=\,1}^{d\left(\mathbf{s}_{i}\right)}\frac{(r_{ij}-\|\mathbf{s}_{i}-\mathbf{s}_{j}\|)^{2}}{\sigma^{2}}.
\end{aligned}
\end{equation}

If solved appropriately, the global MLE can reach the CRLB because
it is consistent with the likelihood function in formula \eqref{eq:=007EA7=008054=004F3C=007136=0051FD=006570}.
However, there are several issues that prevent its application in
MANETs. Firstly, the acquisition of global information inevitably
increases network overhead. Additionally, as the node level increases,
the complexity and dimension of this global information also increase,
leading to increased complexity of the MLE.

In the this subsection, we present the derivation of the proposed
two-step TOA method for network nodes at level 2 or above and conclude
it in Algorithm \ref{alg:two step TOA static}. Our method solely
relies on the estimated positions and distance information of dynamic
anchors, along with their CRLBs. These local information can be transmitted
through single-hop connections, significantly reducing network overhead
and algorithm complexity.

To illustrate our designed algorithm, we take the localization of
level-k node $\mathbf{p}^{k}$ as an example, where $k\geqslant2$.
The total number of dynamic anchors that can be used for location
is $N$. The $i^{th}$ dynamic anchor coordinates and CRLBs are $\mathbf{s}_{i}$
and $\sigma_{ci}^{2}$, respectively. If the $i^{th}$ dynamic anchor
is not a base anchor, we can solve the CRLB by formula \eqref{eq:TOACRLB};
otherwise $\sigma_{ci}^{2}=\delta^{2}$, when the dynamic anchor is
level-0 node. Thus, we can redefine the MLE objective function for
$\mathbf{p}^{k}$ and $\mathbf{s}_{i}$ based on the CRLBs as follows
\begin{equation}
\min_{\mathbf{p}^{k},\mathbf{s}_{i}}~\sum_{i\,=\,1}^{\begin{smallmatrix}N\end{smallmatrix}}\frac{(r_{i}-\|\mathbf{p}^{k}-\mathbf{s}_{i}\|)^{2}}{\sigma^{2}}+\sum_{i\,=\,1}^{\begin{smallmatrix}N\end{smallmatrix}}\frac{\|\tilde{\mathbf{s}}_{i}-\mathbf{s}_{i}\|^{2}}{\sigma_{ci}^{2}}.\label{eq:level-2 target function}
\end{equation}

Considering that intelligent search algorithms offer higher precision
in solving likelihood estimation problems compared to traditional
mathematical methods, we employ the PSO algorithm to solve this problem.
In order to avoid optimization getting trapped in local optima, it
is necessary to narrow down the search space. Firstly, considering
$\mathbf{s}_{i}$ as a dynamic anchor with its own position information
$\tilde{\mathbf{s}}_{i}$ after the previous level positioning, and
base on it setting the search range. Similarly, for node $\mathbf{p}^{k}$
itself, we introduce the iterative Chan algorithm to obtain the approximate
position to set the initial search range. In comparison to the standard
Chan algorithm used for solving the TOA localization equation, our
iterative Chan algorithm can handle coordinate estimation when the
observed distance errors between nodes remain unknown. We name this
method by iterative Chan or iChan for short.

\begin{algorithm}[tbh]
\caption{Iterative Chan (iChan) Algorithm\label{alg:Iterative TOA Chan}}

\renewcommand{\algorithmicrequire}{\textbf{Input:}} 

\begin{algorithmic}[1]

\REQUIRE~~

Dynamic anchors with errors $\tilde{\mathbf{s}}_{i},i=1,2,...,N$;

Observed distances $r_{i}$ between $\mathbf{s}_{i}$ and target $\mathbf{p}$
;

Parameters: maximum iteration count $iter$ and convergence threshold
$\epsilon$

\ENSURE

The estimated target coordinates $\tilde{\mathbf{p}}$ ;

\STATE Initialize the error vector covariance matrix $\boldsymbol{Q}=I_{N}$;

\STATE Initialize the target coordinates $\hat{\bm{z}}_{0}=\left[0,0\right]^{\text{T}}$;

\FOR{ $n=1:iter$ }

\STATE Use formula \eqref{eq:=008FED=004EE3Chan=007B97=006CD5=007B2C=004E00=006B21=004F30=008BA1}
to get $\hat{\bm{z}}_{m}$;

\STATE Calculate $\Psi^{\prime}$ based on formula \eqref{eq:=007B2C=004E8C=006B21=004F30=008BA1=0052A0=006743=007CFB=006570};

\STATE Perform the second WLS estimation to obtain $\hat{\bm{z}}_{p}$
based on formula \eqref{eq:=008FED=004EE3Chan=007B97=006CD5=007B2C2=006B21=004F30=008BA1}
;

\STATE Update parameter estimate $\hat{\bm{z}}_{n}=\pm\sqrt{\hat{\bm{z}}_{p}}$
;

\STATE Compute covariance matrix $\boldsymbol{Q}$ according to formula
\eqref{eq:=007B2C=004E00=006B21WLS=0052A0=006743=007CFB=006570} to
\eqref{eq:Q=00534F=0065B9=005DEE=0077E9=009635};

\IF{ $\left\Vert \hat{\bm{z}}_{n}-\hat{\bm{z}}_{n-1}\right\Vert \leq\epsilon$
}

\STATE Exit $\textbf{\textbf{for}}$ loop

\ENDIF

\ENDFOR 

\STATE $\tilde{\mathbf{p}}=\hat{\bm{z}}_{n}$

\end{algorithmic}
\end{algorithm}

For the objective function in \eqref{eq:level-2 target function},
we assume $\tilde{\mathbf{s}}_{i}=\mathbf{s}_{i}$ and solely estimate
$\mathbf{p}$ to obtain its approximate estimation. During the first
iteration process, assume that $\bm{e}=(e_{1},e_{2},...e_{N})^{\text{T}}$
is squared error vector between $d_{i}^{2}$ and $\|\mathbf{p}-\mathbf{s}_{i}\|^{2}$,
where $i=1,2,...,N$. 
\begin{equation}
\bm{e}=\left[\begin{array}{c}
d_{1}^{2}-R_{1}\\
d_{2}^{2}-R_{2}\\
\vdots\\
d_{N}^{2}-R_{N}
\end{array}\right]-\begin{bmatrix}-2x_{1} & -2y_{1} & 1\\
-2x_{2} & -2y_{2} & 1\\
\vdots & \vdots & \vdots\\
-2x_{N} & -2y_{N} & 1
\end{bmatrix}\begin{bmatrix}x\\
y\\
x^{2}+y^{2}
\end{bmatrix}\label{eq:=005168=0090E8=008BEF=005DEE=005411=0091CF}
\end{equation}
where $R_{i}=x_{i}^{2}+y_{i}^{2}$. To simplify, we rewritten the
squared error vector in \eqref{eq:=005168=0090E8=008BEF=005DEE=005411=0091CF}
as $\bm{e}=\bm{h}-\boldsymbol{G}\bm{z}_{m}$.

The standard Chan algorithm assumes that the squared error vector
$\bm{e}$ follows a Gaussian distribution, which simplifies the process
of obtaining its covariance matrix. However, due to the independence
of observed distance and anchor position errors, the distribution
of the error vector becomes uncertain. As a result, we first assume
the error vector covariance matrix is an n-dimensional identity matrix
during the initial iteration process. Thus, the first weighted least
squares (WLS) estimate of $\bm{z}$ can be obtained as follows 
\begin{equation}
\hat{\bm{z}}_{m}=(\boldsymbol{G}^{T}\boldsymbol{Q}^{-1}\boldsymbol{G})^{-1}\boldsymbol{G}^{T}\boldsymbol{Q}^{-1}\bm{h}\label{eq:=008FED=004EE3Chan=007B97=006CD5=007B2C=004E00=006B21=004F30=008BA1}
\end{equation}
where $\boldsymbol{Q}=\boldsymbol{I}_{N}$ in the first iteration.
Note that the value of $x^{2}+y^{2}$ in $\bm{z}$ is related to the
coordinates of the target. By utilizing the initial estimated position
$\hat{\bm{z}}_{m}$, along with the constraint conditions, a updated
system of equations can be formulated for the second WLS estimation.
The relationship between the first estimation and its true value is
as follows
\begin{equation}
\begin{cases}
\hat{\bm{z}}_{m}(1)=x+e_{1}\\
\hat{\bm{z}}_{m}(2)=y+e_{2}\\
\hat{\bm{z}}_{m}(3)=x^{2}+y^{2}+e_{3}
\end{cases}\label{eq:=007B2C=004E8C=006B21WLS=00521D=0059CB=008BEF=005DEE}
\end{equation}
while $e_{1}$, $e_{2}$ and $e_{3}$ denote estimated errors. By
squaring both $e_{1}$ and $e_{2}$ , the system of equations \eqref{eq:=007B2C=004E8C=006B21WLS=00521D=0059CB=008BEF=005DEE}
can be transformed into the following system
\begin{equation}
\begin{cases}
2xe_{1}+{e_{1}}^{2}=\hat{\bm{z}}_{m}^{2}(1)-x^{2}\\
2ye_{2}+{e_{2}}^{2}=\hat{\bm{z}}_{m}^{2}(2)-y^{2}\\
e_{3}=\hat{\bm{z}}_{m}(3)-\left(x^{2}+y^{2}\right)
\end{cases}\label{eq:=007B2C=004E8C=006B21WLS=008BEF=005DEE=005411=0091CF}
\end{equation}
The left-hand side of this system represents the errors. The equations
\eqref{eq:=007B2C=004E8C=006B21WLS=008BEF=005DEE=005411=0091CF} can
be simplified as
\begin{equation}
\bm{\psi}^{\prime}=\bm{h}^{\prime}-\boldsymbol{G}_{m}^{\prime}\bm{z}_{p}
\end{equation}
where $\text{\ensuremath{\bm{h}}}^{\prime}=\left[\begin{array}{c}
\hat{\bm{z}}_{m}^{2}(1)\\
\hat{\bm{z}}_{m}^{2}(2)\\
\hat{\bm{z}}_{m}(3)
\end{array}\right]$, $\boldsymbol{G}\mathbf{^{\prime}}=\left[\begin{array}{cc}
1 & 0\\
0 & 1\\
1 & 1
\end{array}\right]$ and $\bm{z}_{p}=\left[\begin{array}{c}
x^{2}\\
y^{2}
\end{array}\right]$. $\bm{\psi}^{\prime}$ is error vector of $\bm{z}_{p}$ and can be
transformed into
\begin{equation}
\begin{cases}
\psi_{1}^{\prime}=2xe_{1}+e_{1}^{2}\approx2xe_{1}\\
\psi_{2}^{\prime}=2ye_{2}+e_{2}^{2}\approx2ye_{2}\\
\psi_{3}^{\prime}=e_{3}
\end{cases}\label{eq:=008BEF=005DEEe_ij}
\end{equation}

In order to determine the weights for the second WLS estimate, the
covariance matrix of the first WLS estimate is required, and this
can be achieved using the perturbation method\cite{TOAchan,YU20062153}.
The perturbation of $\bm{z}_{m}$ can be approximated as follows
\begin{equation}
\Delta\bm{z}_{m}\approx(\boldsymbol{G}^{\text{T}}\boldsymbol{Q}^{-1}\boldsymbol{G})^{-1}\boldsymbol{G}^{\text{T}}\boldsymbol{Q}^{-1}\delta
\end{equation}
\begin{equation}
\text{cov}(\bm{z}_{m})=\mathbb{E}[\Delta\bm{z}_{m}\Delta\bm{z}_{m}^{T}]=(\boldsymbol{G}^{\text{T}}\boldsymbol{Q}^{-1}\boldsymbol{G})^{-1}
\end{equation}
Subsequently, the covariance matrix of $\bm{\psi}^{\prime}$ is given
by 
\begin{equation}
\Psi^{\prime}=\mathbb{E}[\bm{\psi}^{\prime}\bm{\psi}^{\prime T}]=4\boldsymbol{B}^{\prime}\text{cov}(\bm{z}_{m})\boldsymbol{B}^{\prime}\label{eq:=007B2C=004E8C=006B21=004F30=008BA1=0052A0=006743=007CFB=006570}
\end{equation}
where $\boldsymbol{B}^{\prime}=\text{diag}(x,y,0.5)$. Consequently,
the estimated value of $Z_{p}$ can be expressed as 
\begin{equation}
\hat{\bm{z}}_{p}=(\boldsymbol{G}^{\prime T}\Psi^{\prime-1}\boldsymbol{G}^{\prime})^{-1}\boldsymbol{G}^{\prime T}\Psi^{\prime-1}\bm{h}^{\prime}\label{eq:=008FED=004EE3Chan=007B97=006CD5=007B2C2=006B21=004F30=008BA1}
\end{equation}

Since $\boldsymbol{B}^{\prime}$ involves the true position of the
target, it can be initially substituted with the estimated values
$\hat{\bm{z}}_{m}(1)$ and $\hat{\bm{z}}_{m}(2)$ obtained from the
first WLS solving process. By performing two consecutive WLS computations,
the localization result of the target is obtained as follows 
\begin{equation}
\hat{\bm{z}}=\pm\sqrt{\hat{\bm{z}}_{p}}\label{eq:=007B2C=004E8C=006B21WLS=007ED3=00679C}
\end{equation}
Next, we utilize $\hat{\bm{z}}$ to optimize the error covariance
matrix for the first WLS. Based on formula \eqref{eq:=008BEF=005DEEe_ij}
and $\delta_{i}\ll d_{i}$, we can derive the estimation error for
the first WLS.
\begin{equation}
e_{i}=2\|\mathbf{p}-\mathbf{s}_{i}\|\xi_{i}+\xi_{i}^{2}\approx2\|\mathbf{p}-\mathbf{s}_{i}\|\xi_{i}
\end{equation}
where $\xi_{i}=d_{i}-\|\mathbf{p}-\mathbf{s}_{i}\|$. Thus, the squared
error vector $\bm{e}$ can be written as
\begin{equation}
\bm{e}=2\boldsymbol{B}\bm{\xi}
\end{equation}

Because $B=\text{diag}(\|\mathbf{p}-\mathbf{s}_{i}\|),i=1,2,...,N$
is the set of true distances between $\mathbf{p}$ and $\mathbf{s}_{i}$,
we substitute $d_{i}$ in place of it. Therefore, the covariance matrix
$\Psi$ of the error vector $\bm{e}$ can be given by 
\begin{equation}
\Psi=\mathbb{E}(\bm{e}\bm{e}^{T})=4\boldsymbol{B}\mathbb{E}(\delta\delta^{T})\boldsymbol{B}=4\boldsymbol{B}\boldsymbol{Q}\boldsymbol{B}\label{eq:=007B2C=004E00=006B21WLS=0052A0=006743=007CFB=006570}
\end{equation}
where $\mathbb{\bm{\xi}}=[\xi_{1},\xi_{2},...,\xi_{N}]^{T}$ and $\boldsymbol{Q}$
is the covariance matrix of the distance error vector. Considering
the coordinates estimated as $\hat{\bm{z}}$ in the second WLS estimation,
we can optimize $\boldsymbol{Q}$ as follows
\begin{equation}
\boldsymbol{Q}=\mathrm{\text{diag}}(\xi_{1}^{2},\xi_{2}^{2},...,\xi_{N}^{2})\label{eq:Q=00534F=0065B9=005DEE=0077E9=009635}
\end{equation}
where 
\begin{equation}
\xi_{i}=r_{i}-\left\Vert \hat{\bm{z}}-\mathbf{s}_{i}\right\Vert .
\end{equation}

Utilizing the estimated coordinates in formula \eqref{eq:=007B2C=004E8C=006B21WLS=007ED3=00679C},
we can optimize the covariance matrix of distance errors for the next
iteration. In summary, each iteration consists of two steps: estimating
coordinates using two consecutive WLS computations and updating the
covariance matrix of distance errors once. This process continues
until either the maximum iteration count $iter$ is reached or the
estimated coordinates from two consecutive iterations satisfy the
convergence criterion $\epsilon$, indicating algorithm convergence.
The iChan algorithm is concluded in Algorithm \ref{alg:Iterative TOA Chan}.

After obtaining the estimation $\tilde{\mathbf{p}}$ and knowing the
approximate coordinates $\tilde{\mathbf{s}}_{i}$ of the dynamic anchors,
we can set its neighborhood as the search domain to solve \eqref{eq:level-2 target function}.
In order to fully leverage the positional information and CRLB of
neighboring nodes, we propose an advanced two-step TOA localization
algorithm based on CRLBs. Define $B(\mathbf{\bm{s}},r)$ to represent
a circular neighborhood with its center at coordinate point $\mathbf{\bm{s}}$
and a radius of $r$. Thus the MLE problem in \eqref{eq:level-2 target function}
can be written as
\begin{equation}
\begin{aligned}\min_{\substack{\mathbf{p}^{k},\,\mathbf{s}_{i}}
} & \;~~\sum_{i\,=\,1}^{\begin{smallmatrix}N\end{smallmatrix}}\frac{(r_{i}-\|\mathbf{p}^{k}-\mathbf{s}_{i}\|)^{2}}{\sigma^{2}}+\sum_{i\,=\,1}^{\begin{smallmatrix}N\end{smallmatrix}}\frac{\|\tilde{\mathbf{s}}_{i}-\mathbf{s}_{i}\|^{2}}{\sigma_{ci}^{2}}\\
\text{s.t.} & \;\;\;\;\begin{aligned}\mathbf{s}_{i}\in B(\tilde{\mathbf{s}}_{i},r_{s})\;\;\\
\mathbf{p}^{k}\in B(\tilde{\mathbf{p}}^{k},r_{p})
\end{aligned}
\end{aligned}
\label{eq:level-2 first estimate}
\end{equation}

In Equation \eqref{eq:level-2 first estimate}, it is noteworthy that
while solving for $\mathbf{p}$, estimates of its neighboring dynamic
anchors $\mathbf{s}_{i}$ are simultaneously derived by utilizing
lower-level neighbors. After employing the PSO algorithm, the updated
dynamic anchors are denoted as $\tilde{\mathbf{s}}_{i}^{\prime}$. 
\begin{rem}
\label{rem:update CRLB}The updated estimation accuracy of dynamic
anchors is higher than that of the initial estimation. Intuitively,
$\tilde{\mathbf{s}}_{i}^{\prime}$, on top of $\tilde{\mathbf{s}}_{i}$,
integrates distance constraints for serving as dynamic anchors in
localizing higher-level nodes. This integration is depicted by the
first term in Equation \eqref{eq:level-2 first estimate} , resulting
in significantly improved performance. The validation of this enhancement
finds support in the CRLB and can be further proven mathematically
in Appendix \ref{sec:PROOF-OF-REMARK}.
\end{rem}
Given the superior precision of $\tilde{\mathbf{s}}_{i}^{\prime}$
compared to $\tilde{\mathbf{s}}_{i}$, we employ $\tilde{\mathbf{s}}_{i}^{\prime}$
as the estimated dynamic anchors for the secondary estimation of $\mathbf{p}$.
The objective function is as follows
\begin{equation}
\begin{aligned}\min_{\substack{\mathbf{p}^{k},\,\mathbf{s}_{i}}
} & \;\;~\sum_{i\,=\,1}^{\begin{smallmatrix}N\end{smallmatrix}}\frac{(d_{i}-\|\mathbf{p}^{k}-\mathbf{s}_{i}\|)^{2}}{\sigma^{2}}+\sum_{i\,=\,1}^{\begin{smallmatrix}N\end{smallmatrix}}\frac{\|\tilde{\mathbf{s}}_{i}^{\prime}-\mathbf{s}_{i}\|^{2}}{(\sigma_{ci}^{2})^{\prime}}\\
\text{s.t.} & \;\;\;\;\begin{aligned}\mathbf{s}_{i}\in B(\tilde{\mathbf{s}}_{i},r_{s})\;\;\\
\mathbf{p}^{k}\in B(\tilde{\mathbf{p}}^{k},r_{p})
\end{aligned}
\end{aligned}
\label{eq:level-2 second estimate}
\end{equation}
where the first term aligns with Equation \eqref{eq:level-2 first estimate}.
On the other hand, $(\sigma_{ci}^{2})^{\prime}$ represents a modification
of $\sigma_{ci}^{2}$. This modification incorporates information
from the target $\mathbf{p}^{k}$ into the FIM, resulting in a reduced
CRLB utilized for the subsequent estimation. Ultimately, this problem
remains amenable to resolution using the PSO algorithm, facilitating
the attainment of the final estimation outcome for the subsequent
step.

\begin{algorithm}[tbh]
\caption{Two-step TOA Localization Algorithm Based on CRLB\label{alg:two step TOA static}}

\renewcommand{\algorithmicrequire}{\textbf{Input:}} 

\begin{algorithmic}[1]

\REQUIRE~~

Initial dynamic anchor estimate $\tilde{\mathbf{s}}_{i}$, $i=1,2,...,N$;

Initial CRLBs for dynamic anchors $\sigma_{ci}^{2}$, $i=1,2,...,N$;

Observed distances $d_{i}$ between $\mathbf{s}_{i}$ and target $\mathbf{p}$;

\ENSURE

The estimated coordinates of target nodes $\mathbf{\hat{p}}$;

\STATE Use iChan method in Algorithm \ref{alg:Iterative TOA Chan}
to obtain approximate estimation $\tilde{\mathbf{p}}$ for the target
node $\mathbf{p}$;

\STATE Form search neighborhood $B(\tilde{\mathbf{s}}_{i},r_{s})$
and $B(\tilde{\mathbf{p}},r_{p})$;

\STATE Solve problem in \eqref{eq:level-2 first estimate} using
the PSO method to update the dynamic anchors $\tilde{\mathbf{s}}_{i}^{\prime}$,
$i=1,2,...,N$;

\STATE Update CRLBs $(\sigma_{ci}^{2})^{\prime}$ for dynamic anchor
errors by employing $\tilde{\mathbf{s}}_{i}^{\prime}$;

\STATE Solve problem in \eqref{eq:level-2 second estimate} using
the PSO method to get the final estimation $\mathbf{\hat{p}}$;

\end{algorithmic}
\end{algorithm}

\subsection{Robust TOA Algorithm for Mobile Environments}

In MANETs, node mobility stands as a formidable challenge for localization
due to its profound impact on network topology and anchor point selection.
Taking into account that each network node possesses memory and storage
capabilities, we have enhanced the static algorithm's performance
by incorporating the previous moment's location information into the
existing algorithm, making our method more robust in MANETs.

Given the pronounced dynamics of the MANET, both base anchors and
other blind nodes adhere to a mobility model that merges the Random
Waypoint Mobility Model (RWM) and Random Direction (RD) model. This
mobility model computes node movement speed as a vector summation
of average and random speeds. Consequently, we represent the coordinates
of a dynamic node within a MANET at a specific sampling time as follows
\begin{align}
x_{t} & =x+v_{2}\times\cos\left(\theta_{2}\right)\times\varDelta t\nonumber \\
 & =x+(v_{\text{mean }}+v_{n})\cos\left(R_{\theta}(t)\right)\varDelta t\nonumber \\
y_{t} & =y+v_{2}\times\sin(\theta_{2})\times\varDelta t\nonumber \\
 & =y+(v_{\text{mean }}+v_{n})\sin\left(R_{\theta}(t)\right)\varDelta t\label{eq:mobility model}
\end{align}
where $\mathbf{p}=\left(x,y\right)^{\text{T}}$ and $\mathbf{p}_{t}=\left(x_{t},y_{t}\right)^{\text{T}}$
are the initial and post-movement coordinates of dynamic nodes, respectively. 

Consider a target node $\mathbf{p}$ at sampling time $t_{1}$, denoted
as $\mathbf{p}_{t1}$, and at a subsequent time $t_{2}$, denoted
as $\mathbf{p}_{t2}$. Assuming $t_{2}>t_{1}$, we can obtain the
past-time estimation $\hat{\mathbf{p}}_{t1}$ at next time $t_{2}$.
Our dynamic approach aims to optimize the current computation through
past-time position estimations. It's important to note that estimations
of past-time node positions are reliant on anchor-based localization,
and thus their accuracy remains uncertain. Furthermore, the motion
model of the node involves intricate, random movement patterns. Therefore,
it is not reasonable to impose strict and inflexible constraints directly
in our dynamic method. 

In this subsection, we introduce a penalty term as a soft constraint
by penalizing excessive position changes between two sampling time.
The penalty factor is adjusted to maintain the balance of the previous
estimation rather than imposing hard constraints. Illustrated in Algorithm
\ref{alg:two step TOA static}, the first step aims to refine anchor
coordinates in our original two-step TOA algorithm, as outlined in
the formula \eqref{eq:level-2 first estimate}. This step remains
unchanged in our dynamic algorithm. However, in the context of the
second step, alongside optimizing anchor coordinates, our dynamic
algorithm introduces mobility constraints. Therefore, the optimization
problem presented in the formula \eqref{eq:level-2 second estimate}
can be reformulated as follows
\begin{equation}
\begin{aligned}\min_{\substack{\mathbf{p}_{t2},\,\mathbf{s}_{i}}
} & \;\;\underset{i=1}{\overset{N}{\mathop{\sum}}}\left(\frac{\left(r_{i}-d_{i}\right)^{2}}{\sigma^{2}}+\frac{\|\tilde{\mathbf{s}}_{i}^{\prime}-\mathbf{s}_{i}\|^{2}}{(\sigma_{ci}^{2})^{\prime}}\right)+\eta\left(r_{t1t2}-d_{m}\right)^{2}\\
\text{s.t.} & \begin{aligned} & \;\;\;\mathbf{s}_{i}\in B(\tilde{\mathbf{s}}_{i}^{\prime},r_{i})\\
 & \;\;\;\mathbf{p}_{t2}\in B(\mathbf{\tilde{p}}_{t2},r)\\
 & \;\;\;d_{m}=v_{\text{mean }}\left(t_{2}-t_{1}\right)\\
 & \;\;\;d_{i}=\|\mathbf{p}_{t2}-\mathbf{s}_{i}\|\\
 & \;\;\;r_{t1t2}=\|\mathbf{p}_{t2}-\mathbf{\hat{p}}_{t1}\|
\end{aligned}
\end{aligned}
\label{eq:level-2 second step dynamic}
\end{equation}
where $\eta$ is the penalty factor. When the time interval between
$t_{2}$ and $t_{1}$ is sufficiently small, the motion of node $\mathbf{p}$
can be approximated as linear, and accordingly, we establish the mobility
constraint. Ultimately, the PSO algorithm can then be employed to
address this problem and yield the conclusive estimation outcome for
the second step. The refined two-step TOA localization algorithm tailored
for mobile environments is succinctly outlined in Algorithm \ref{alg:two step TOA dynamic}.

\begin{algorithm}[tbh]
\caption{Two-step TOA localization algorithm for mobile environments\label{alg:two step TOA dynamic}}

\renewcommand{\algorithmicrequire}{\textbf{Input:}} 

\begin{algorithmic}[1]

\REQUIRE~~

Initial dynamic anchor estimate $\tilde{\mathbf{s}}_{i}$, $i=1,2,...,N$;

Initial CRLBs for dynamic anchors $\sigma_{ci}^{2}$, $i=1,2,...,N$;

Observed distances $d_{i}$ between $\mathbf{s}_{i}$ and target $\mathbf{p}_{t2}$;

Previous target estimate $\mathbf{\hat{p}}_{t1}$ at past time $t_{1}$;

\ENSURE

The current coordinates estimate of target nodes $\mathbf{\hat{p}}_{t2}$;

\STATE Use iChan method in Algorithm \ref{alg:Iterative TOA Chan}
to obtain approximate estimation result $\mathbf{\tilde{p}}_{t2}$
for the target node $\mathbf{p}$;

\STATE Form search neighborhood $B(\tilde{\mathbf{s}}_{i},r_{s})$
and $B(\tilde{\mathbf{p}},r_{p})$;

\STATE Solve problem in \eqref{eq:level-2 first estimate} using
PSO method to update the estimation result for dynamic anchors $\tilde{\mathbf{s}}_{i}^{\prime}$,
$i=1,2,...,N$;

\STATE Update CRLBs $(\sigma_{ci}^{2})^{\prime}$ for dynamic anchor
errors by employing $\tilde{\mathbf{s}}_{i}^{\prime}$;

\STATE Solve problem in \eqref{eq:level-2 second step dynamic} using
the PSO method to get the final estimation $\mathbf{\hat{p}}_{t1}$;

\end{algorithmic}
\end{algorithm}

\section{Numerical Simulations\label{sec:Results and system performance}}

In this section, we conduct numerical simulations to assess our proposed
method's robustness within MANETs. Firstly, we estimate the coordinates
of a non-direct node to validate the performance of our approach in
simple scenarios. We employ the CRLB as the benchmark to assess the
accuracy of our estimations. Following this, we simulate a scenario
of a mobile ad hoc network, applying our node localization method
to tackle the challenges posed by highly dynamic environments.

\subsection{Evaluation of Two-step TOA Localization Algorithm Based on CRLB}

In this subsection, we validate our designed TOA algorithm, assessing
its localization accuracy for a single node while benchmarking it
against other established algorithms. Our algorithm specifically caters
to localization concerns in level-2 or above nodes. Thus, we set the
communication radius between nodes at 500 m, establishing the foundation
for hierarchical localization. To simplify, we envision a scenario
comprising nine nodes: four at level 0, four at level 1, and one at
level 2. Assuming an initial presence of four base anchors, each characterized
by Gaussian-distributed position errors, we randomly select the true
coordinates $\mathbf{s}_{i}^{0}$ to be 
\[
\left(134,103\right)^{\text{T}},\left(155,205\right)^{\text{T}},\left(103,220\right)^{\text{T}},\left(35,264\right)^{\text{T}}
\]
However, we only know the coordinates of anchor nodes with additive
Gaussian errors whose standard deviation (represented by $\delta\mathbf{I}_{2}$)
is 3 m for each component. Furthermore, the initial positions of the
level-1 nodes $\mathbf{p}_{i}^{1}$ are established as follows 
\[
\left(431,232\right)^{\text{T}},\left(324,577\right)^{\text{T}},\left(200,398\right)^{\text{T}},\left(498,245\right)^{\text{T}}
\]

Considering a level-2 node $\mathbf{p}^{2}$ as the target with an
initial position of $\left(600,450\right)^{\text{T}}$, the observed
distances between each connected pair of nodes follow a Gaussian distribution
with a variance of $\sigma$. In the first step of our method, we
aim to obtain more accurate estimates of neighbor dynamic anchors.
For example, $\mathbf{p}_{1}^{1}=\left(431,232\right)^{\text{T}}$
is a level-1 node, it can be estimated by level-0 nodes as well as
serves as a dynamic anchor point to locate level-2 node $\mathbf{p}^{2}$.
We compute the Root Mean Squared Error (RMSE) for the positioning
outcomes through multiple simulations, which is determined by the
number of simulations $N$ and can be expressed as 
\begin{equation}
\text{ RMSE=}\sqrt{\frac{1}{N}\underset{i=1}{\overset{N}{\mathop{\sum}}}((\hat{x}_{i}-x)^{2}+(\hat{y}_{i}-y)^{2})}
\end{equation}

\begin{figure}[tbh]
\begin{centering}
\includegraphics[width=8cm]{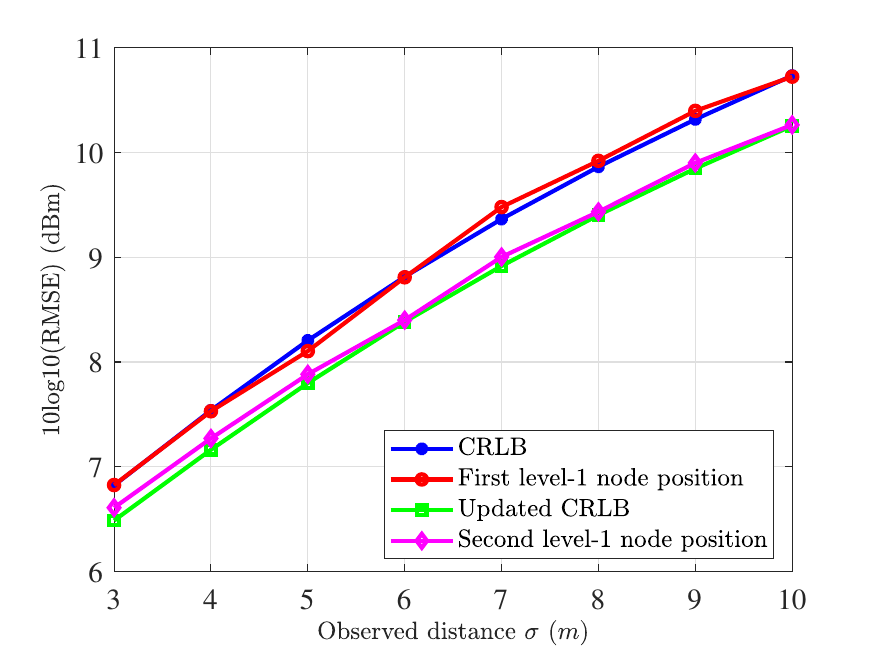}
\par\end{centering}
\caption{Illustration of one level-1 node positioning error in different distance
errors\label{fig:level-1 node different step}}
\end{figure}

In Fig. \ref{fig:level-1 node different step}, the depiction showcases
the position errors of node $\mathbf{p}_{1}^{1}$ based on two independent
estimates affected by different distance errors. In the initial estimation,
node $\mathbf{p}_{1}^{1}$ solely serves as a target point, utilizing
level-0 nodes for localization, achieving accuracy aligned with the
initial CRLB. On the other hand, if this node functions as a dynamic
anchor for the level-2 node $\mathbf{p}^{2}$, the resultant CRLB
is lower than the initial one. For this reason, a subsequent estimation
for this level-1 node is pursued. In accordance with the first step
of our method, this supplementary estimation for the level-1 node
approaches the updated CRLB accuracy. Examining the outcomes across
varying distance errors in Fig. \ref{fig:level-1 node different step},
when the distance measurement errors among the network nodes escalate,
the second positional outcome for the level-1 node consistently outperforms
its initial estimation results.

\begin{figure}[tbh]
\begin{centering}
\includegraphics[width=8cm]{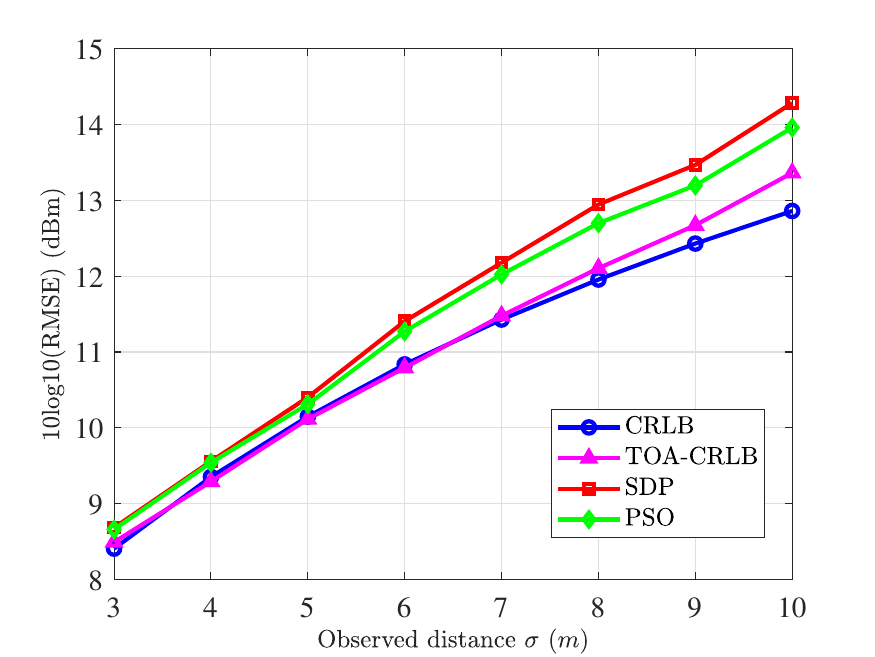}
\par\end{centering}
\caption{Impact of the distance errors on the RMSE performance of level-2 node
positioning\label{fig:level-2 node different step}}
\end{figure}

In Fig. \ref{fig:level-2 node different step}, the accuracy of our
proposed algorithm for locating a level-2 node is verified and the
robustness of the algorithm is examined by setting diverse observed
distance measurement errors. Consistently, we maintain a fixed standard
deviation of 3 m for the base anchor position error. To our knowledge,
since there are fewer algorithms that use TOA multilevel localization
within MANET scenarios, our comparative analysis involves contrasting
our algorithm with other direct TOA localization methods, particularly
in scenarios where anchor points exhibit errors.

The simulation results demonstrate that our algorithm achieves remarkable
proximity to the CRLB for localizing a level-2 node that cannot be
directly located by base anchors when the distance error remains small.
In Fig.\ref{fig:level-2 node different step}, our two-step TOA-CRLB
localization algorithm, closely approaches the theoretical error limit
when the standard deviation of observed distance errors measures 8
m or less. However, as the observed distance error continues to increase,
our algorithm's performance starts to deteriorate gradually.

Comparatively, in contrast to conventional mathematical solutions
represented by the SDP algorithm and optimization algorithms like
PSO, neither category can attain the CRLB even in scenarios with minimal
observed distance noise. Moreover, when confronted with substantial
observed distance errors, both these algorithm types degrade rapidly.
Notably, when the standard deviation of the distance error surpasses
5 m, degradation becomes more pronounced, reaching a critical point
above 8 m. Conversely, our algorithm maintains a higher level of accuracy,
exhibiting superior resilience under challenging conditions with large
observed distance errors.

\begin{figure}[tbh]
\begin{centering}
\includegraphics[width=8cm]{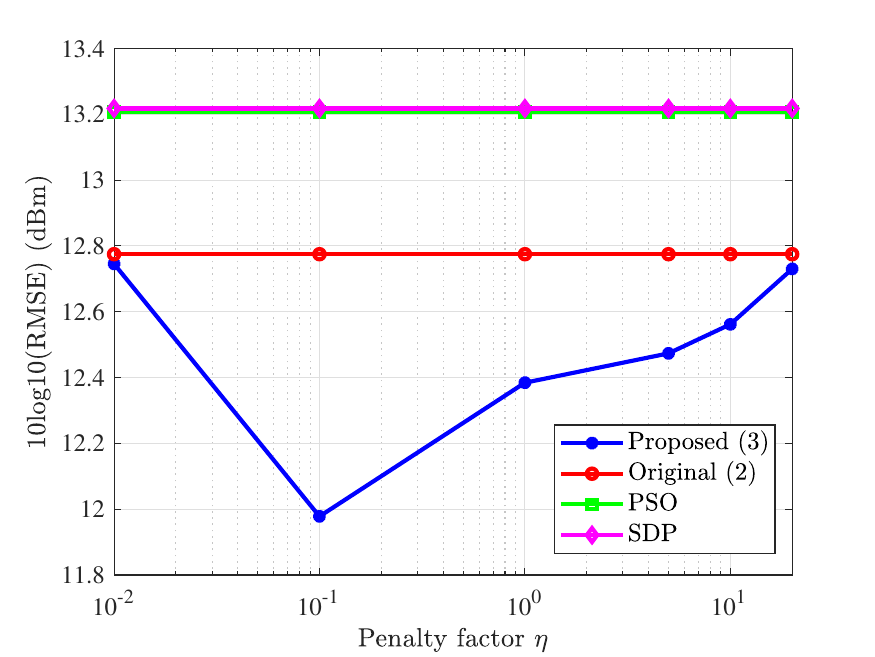}
\par\end{centering}
\caption{RMSE performance of the proposed method in \ref{alg:two step TOA dynamic}
under various penalty factors $\eta$\label{fig:level-2 node penalty factors}}
\end{figure}

To assess the robustness of our dynamic algorithm in a mobile environment,
we implement the mobility models described in the formula \eqref{eq:mobility model}
across all nodes. The average velocity and time intervals are set
at 10 m/s and 1 second, respectively. The standard deviations for
base anchors and observed distance errors are 3 m and 5 m, correspondingly.
In Fig. \ref{fig:level-2 node penalty factors}, we evaluate the performance
of the proposed algorithm \ref{alg:two step TOA dynamic} concerning
the selection of penalty factor $\eta$. Additionally, we compare
its performance with the initial algorithm outlined in Algorithm \ref{alg:two step TOA static},
alongside other existing TOA localization algorithms.

The penalty factor is set to vary from $10^{-2}$ to $20$. From Fig.
\ref{fig:level-2 node penalty factors}, it's evident that both the
PSO and SDP algorithms maintain a considerable distance from our original
static method in Algorithm \ref{alg:two step TOA static}, which is
consistent with results in Fig. \ref{fig:level-1 node different step}.
Further simulations reveal that the RMSE of our dynamic method in
Algorithm \ref{alg:two step TOA dynamic} stabilizes at the 12 dBm
when penalty factor $\eta$ is set to $10^{-1}$. In comparison to
our static method, it offers an improvement of around 0.8 dBm. Moreover,
when contrasted with other algorithms, the accuracy gain is notably
more substantial. Therefore, we opt to set the penalty factor to $10^{-1}$
for the subsequent validation processes.

\subsection{2D Localization System in a MANET Environment}

\begin{figure*}[tbh]
\begin{centering}
\subfloat{\centering{}\includegraphics[width=6.1cm]{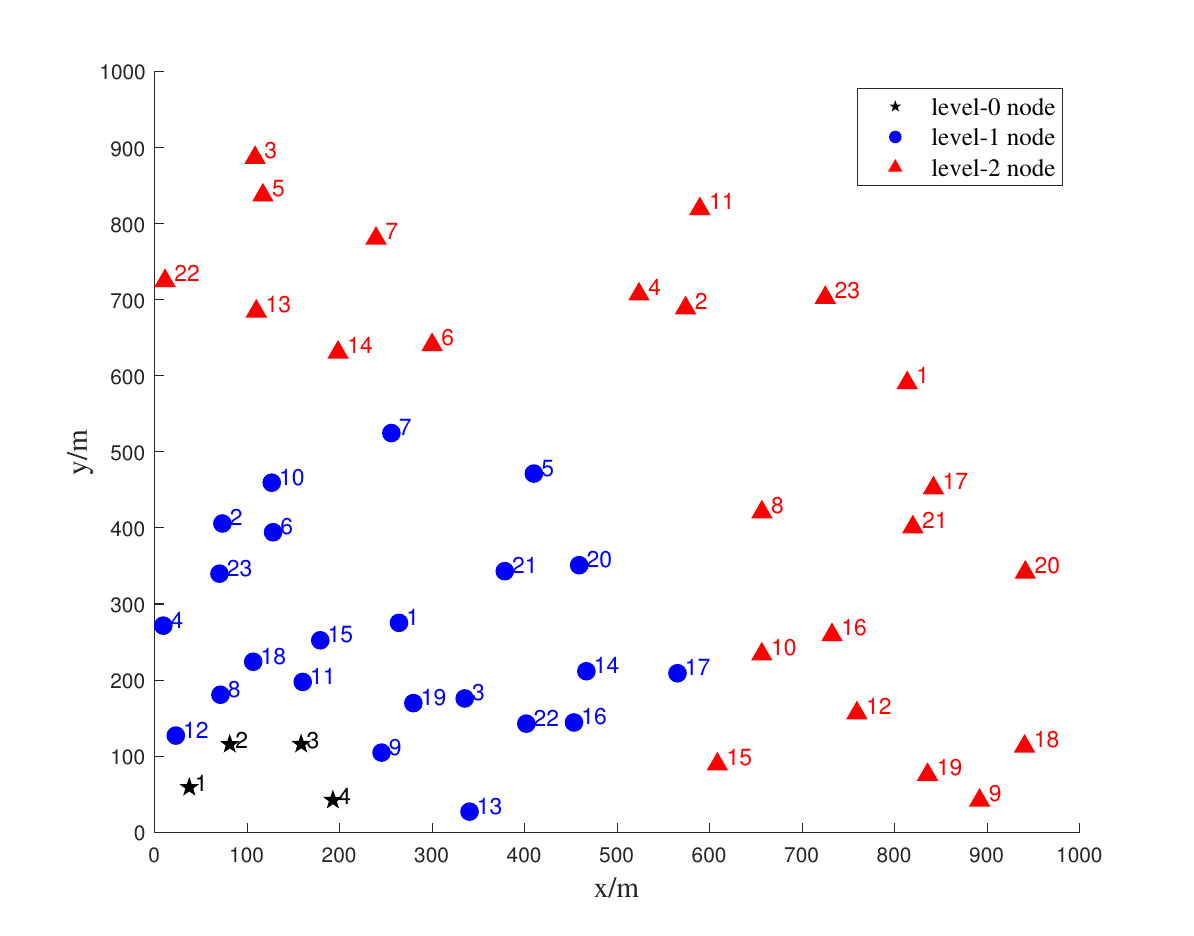}}%
\subfloat{\centering{}\includegraphics[width=6.1cm]{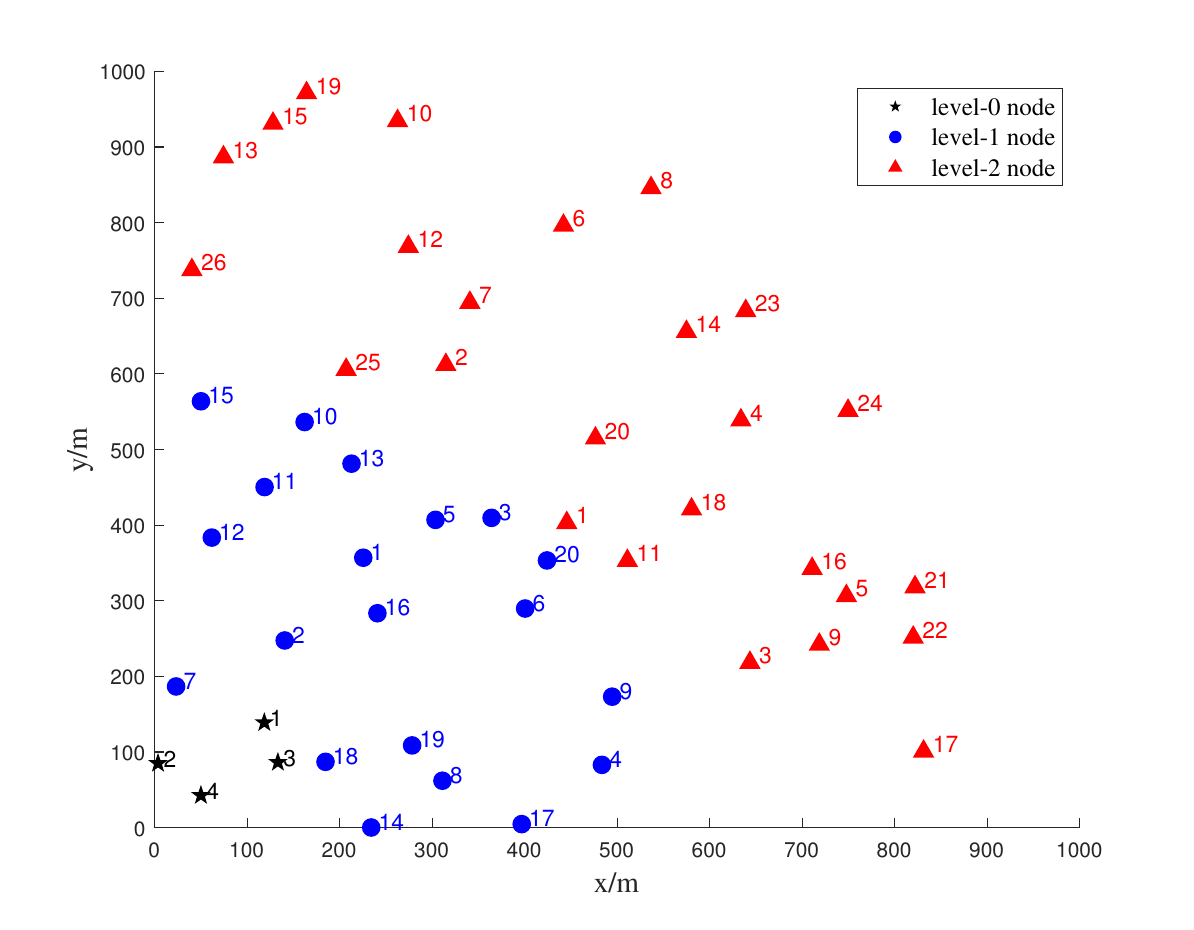}}%
\subfloat{\centering{}\includegraphics[width=6.1cm]{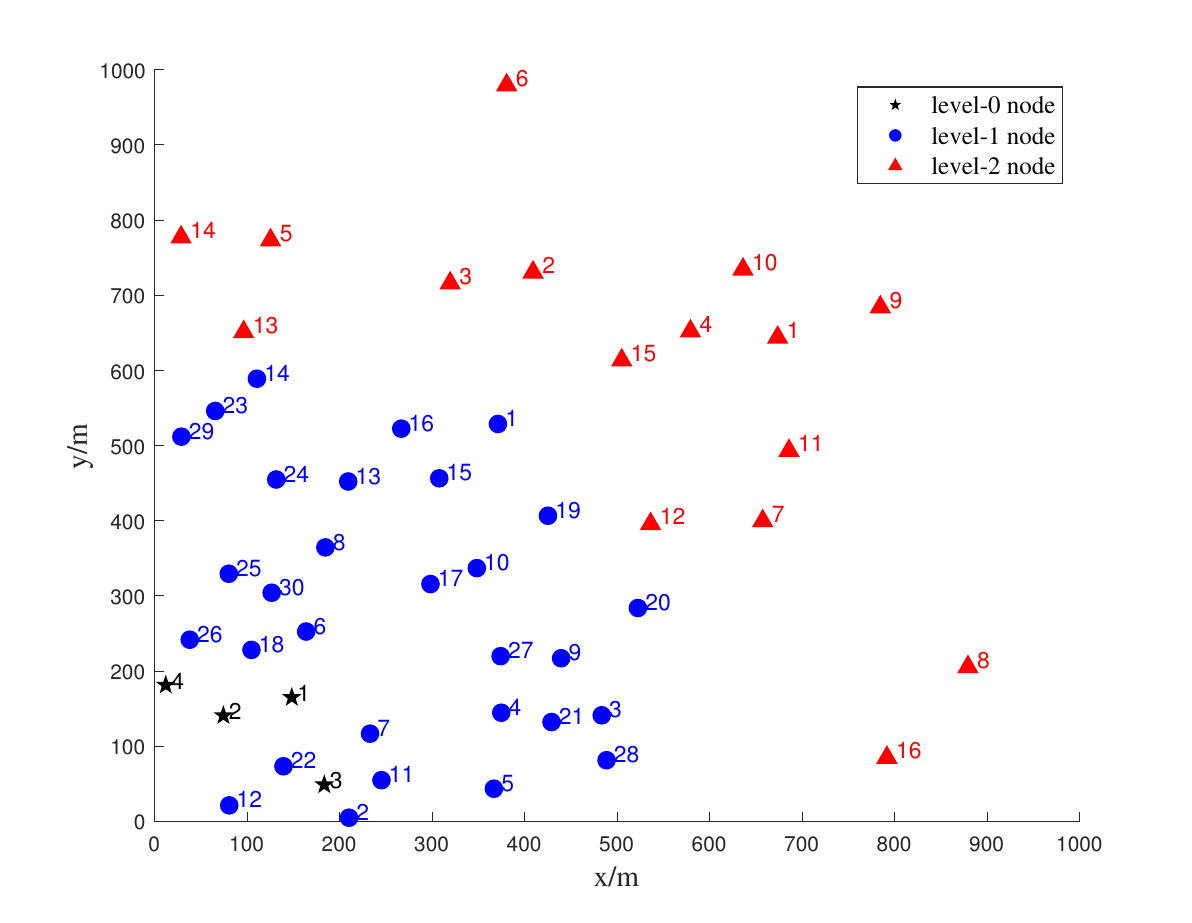}}
\par\end{centering}
\caption{Network node placement topology hierarchy at different time (a) movement
time $T=5s$; (b) movement time $T=15s$; (c) movement time $T=40s$.\label{fig:Initial-node-topology}}
\end{figure*}

For the practical realization of our designed TOA methodology, we
create a multi-hop MANET scenario and perform simulations within a
2D localization system encompassing a square area of $1000m\times1000m$,
depicted in Fig. \ref{fig:Initial-node-topology}. The network consists
of 50 nodes, where 4 nodes are designated as inaccurate base anchors
at level 0, while the highest-tier nodes in the network are established
as level-2 nodes. All network nodes exhibit dynamic behavior, maintaining
an average movement speed from 10 m/s to 50 m/s.

Apart from the base nodes, functioning as level-0 nodes within the
localization network, the hierarchical levels of other network nodes
necessitate dynamic determination based on the quantity and levels
of their neighboring nodes. These levels undergo real-time modifications
within MANETs. Fig. \ref{fig:Initial-node-topology} visually illustrates
the evolving localization hierarchy across different time frames,
showcasing dynamic fluctuations in the numbers of level-1 and level-2
nodes. Moreover, specific nodes adapt their levels and localization
strategies in response to alterations in network topology. Due to
the dynamic and complex nature of these environments, localizing the
remaining nodes becomes more challenging.

\subsection{Performance Evaluation of Level-2 Node Localization in MANETs}

In this subsection, we showcase the practical application of our proposed
algorithm for localizing level-2 nodes within MANETs, and compare
with other wireless localization algorithms based on TOA measurements.
The parameters of the simulation scene are given in Table \ref{tab:SIMULATION-PARAMETERS}.

\begin{table}[tbh]
\begin{raggedright}
\caption{SIMULATION PARAMETERS\label{tab:SIMULATION-PARAMETERS}}
\par\end{raggedright}
\centering{}%
\begin{tabular}{cccc}
\hline 
parameter &
value &
parameter &
value\tabularnewline
\hline 
node number &
50 &
base node number &
4\tabularnewline
transmission radius $r$ &
500m &
average speed $v_{\text{mean }}$ &
10-50m/s\tabularnewline
random speed $v_{n}$ &
0-5m/s &
motion angle $R_{\theta}(t)$ &
0-360�\tabularnewline
movement time $T$ &
20s &
sampling interval $\Delta t$ &
0.2s\tabularnewline
observed distance SD $\sigma$ &
3-10m &
anchor position SD $\delta$ &
3m\tabularnewline
\hline 
\end{tabular}
\end{table}

In our simulation scenario setup, the highest level for MANET nodes
is 2, restricting direct connections with imprecise base anchors.
Additionally, it is possible to expand node localization to higher
levels beyond this limit. For level-1 nodes within MANETs, determining
node positions still relies on error-prone anchor-based methods, considering
the variability in available anchor locations and the varying numbers
of neighboring base anchors. This approach mirrors the research presented
in \cite{TOAsynchronization,SDPYANG,SDPzou,SDPmekon}. Consequently,
our algorithm primarily concentrates on resolving the localization
challenges for nodes that are not directly located by level-0 nodes.

In the case of level-2 nodes, the number of these nodes varies over
time, and any level-2 node can transition to a level-1 node due to
random movement patterns. Therefore, analyzing the alterations of
a single node is inadequate within practical scenario. Instead, it
is imperative to compare the average RMSE of all level-2 nodes to
validate the robustness of our algorithm, which is defined by $\varepsilon_{level2}=\frac{1}{n_{2}}\text{\ensuremath{\underset{i=1}{\overset{n_{2}}{\mathop{\sum}}}}RMSE\ensuremath{\left(\mathbf{p}_{i}^{2}\right)}}$.

\begin{figure}[tbh]
\begin{centering}
\includegraphics[width=8cm]{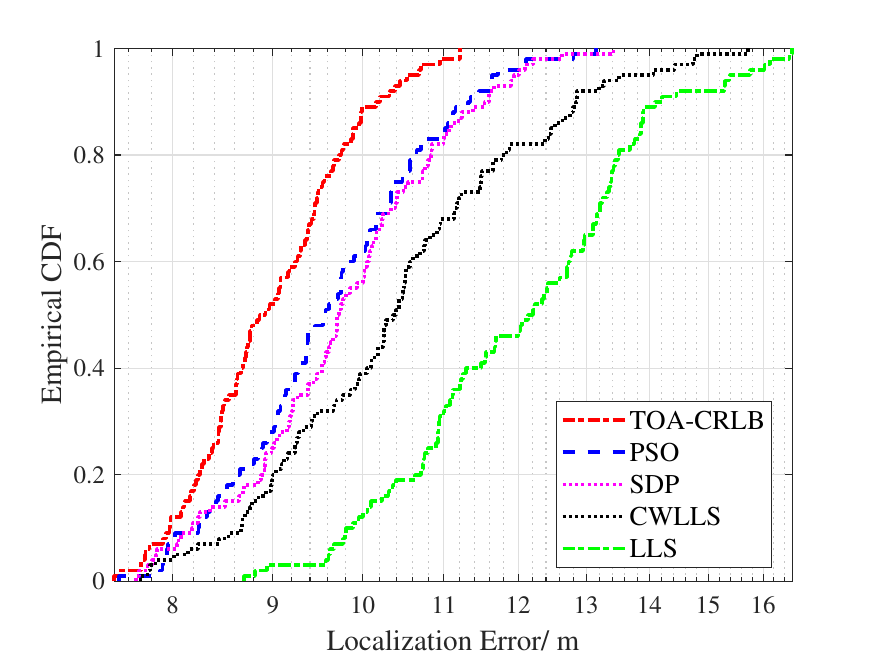}
\par\end{centering}
\caption{CDF curve of average error in positioning level-2 nodes\label{fig:Level-2-node CDF}}
\end{figure}

Fig. \ref{fig:Level-2-node CDF} illustrates the cumulative distribution
function (CDF) of $\varepsilon_{level2}$ across each sampling instance,
with node average speed set at 20 m/s and observed distance standard
deviation at 5 m. Our approach demonstrates remarkable accuracy in
positioning estimation, with 90\% of the average RMSE for all level-2
nodes held to within 10.2 m. 

Furthermore, it demonstrates consistent positioning stability by maintaining
an error margin of 3.71 m for all level-2 nodes. This greatly reduces
the probability of target estimations with significant errors. Specifically,
our median error reaches 8.85 m, which is in contrast to the median
errors of 9.55 m, 9.71 m, 10.40 m, and 12.54 m observed in PSO\cite{TOApso},
SDP\cite{SDPmekon}, CWLLS\cite{TOAchan} and LLS\cite{TOALS} methods,
respectively. This superiority stems from its robustness against abrupt
network topology changes and its adept utilization of reliable neighbor
node positions.

\begin{figure}[tbh]
\begin{centering}
\includegraphics[width=8cm]{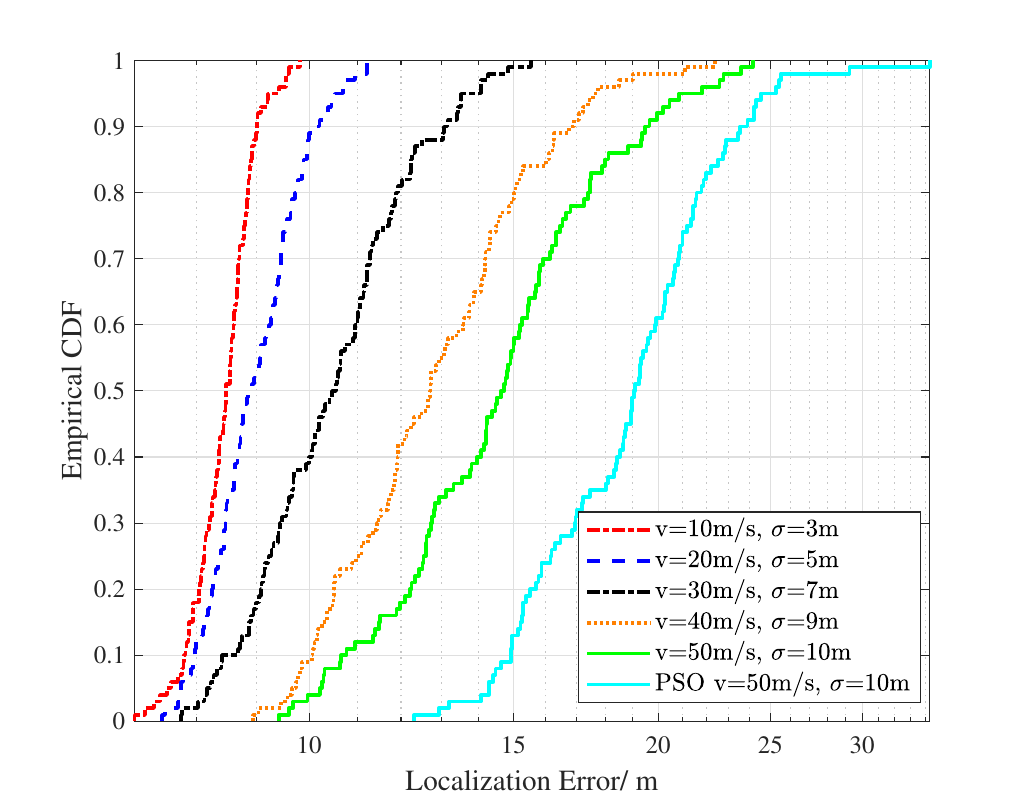}
\par\end{centering}
\caption{Illustration of level-2 node positioning error under different velocities
\label{fig:Level-2-node CDF-1}}
\end{figure}

In order to further verify the robustness of our method, we control
the average speed of network nodes between 10 m/s and 50 m/s to ensure
that the simulation results are consistent across a range of velocities.
For the same sampling interval, besides the increased average movement
speed, the standard deviation of observed distance errors $\sigma$
has also increased from 3 m to 10 m. As illustrated in Fig. \ref{fig:Level-2-node CDF-1},
our method exhibits higher accuracy when the mobility speed and distance
errors are low. When the mobility speed is below 20 m/s, our algorithm
can achieve an accuracy of 10 m which meet the node positioning accuracy
requirements for most networks under dynamic conditions\cite{localization6G}.
Moreover, it maintains accuracy even when network node velocities
undergo fluctuations. Even with a node average speed of 50 m/s, our
algorithm still achieves a median error of 14.73 m for locating level-2
nodes. Notably, our method consistently yields smaller average errors
compared to the utilization of the PSO algorithm, whose median error
stands at 19.07 m. By using the CRLBs, the incorporation of two-step
positioning outcomes not only enhances positioning accuracy but also
mitigates the abrupt oscillations in dynamic anchor position errors
resulting from the dynamic movements of nodes within the MANET.

\section{Conclusions}

In this paper, we refine the TOA localization algorithm tailored for
optimal deployment in MANETs. Firstly, we derive the CRLB as an evaluation
benchmark and a guiding principle for our algorithm's design, specifically
catering to nodes position errors within a hierarchical MANET scenario.
Secondly, we introduce a novel two-step TOA-CRLB localization approach
to mitigate the error diffusion in cascaded localization. Leveraging
CRLB insights and neighboring node positions, our method corrects
neighbor node positions in the initial step to derive the final result.
Extensive experimental results attest to the robustness and accuracy
of our method, showcasing its superiority amidst substantial observed
noise and complex dynamic settings compared to other TOA-based techniques.
Future work will focus on integrating TOA with other localization
algorithms to enhance the robustness of positioning accuracy in more
challenging scenarios. Additionally, we aim to extend CRLB applications
to enhance the precision of diverse wireless localization algorithms.

\appendices{}

\section{DERIVATION OF THE FIM\label{sec:DERIVATION FIM}}

We first define a concept where two nodes, $\mathbf{s}_{i}$ and $\mathbf{s}_{j}$,
satisfy the communication radius and possess different hierarchical
levels. We term such nodes \textquotedbl cross-level nodes\textquotedbl ,
where the lower-level node can serve as the dynamic anchor for the
higher-level node. In cascading localization scenarios, we define
$\bm{\theta}=\left(\mathbf{p}^{k},\bm{L}_{k-1},\bm{L}_{k-2},...,\bm{L}_{0}\right)$
as the estimator in the MLE framework. Based on this estimator, we
derive the FIM in our localization system. The log-likelihood function
of the location estimation (ignoring the constant term) is given by
\begin{align}
\text{L}\left(\bm{\theta}\right) & \triangleq\sum_{i\,=\,1}^{d\left(\mathbf{p}^{k}\right)}f_{1}\left(d_{i}\right)+\sum_{i\,=\,1}^{\mathbf{n}_{k-1}}\sum_{j\,=\,1}^{d\left(\mathbf{s}_{i}\right)}f_{2}\left(d_{ij}\right)+\sum_{i\,=\,1}^{\mathbf{n}_{0}}f_{3}\left(\mathbf{s}_{i}^{0}\right)\nonumber \\
f_{1}\left(d_{i}\right) & =-\frac{1}{2\sigma^{2}}\left(r_{i}-\|\mathbf{p}^{k}-\mathbf{s}_{i}\|\right)^{2},\mathbf{s}_{i}\in D\left(\mathbf{p}^{k}\right)\nonumber \\
f_{2}\left(d_{ij}\right) & =-\frac{1}{2\sigma^{2}}\left(r_{ij}-\|\mathbf{s}_{i}-\mathbf{s}_{j}\|\right)^{2},\mathbf{s}_{j}\in D\left(\mathbf{s}_{i}\right)\nonumber \\
f_{3}\left(\mathbf{s}_{i}^{0}\right) & =-\frac{1}{2\delta^{2}}\|\mathbf{s}_{i}^{0}-\tilde{\mathbf{s}}_{i}^{0}\|^{2},\mathbf{s}_{i}\in\bm{L}_{0}.
\end{align}
where $d_{i}=\|\mathbf{p}^{k}-\mathbf{s}_{i}\|$ and $d_{ij}=\|\mathbf{s}_{i}-\mathbf{s}_{j}\|$.
Substituting $\mathbb{E}(r_{i})=d_{i}$, $\mathbb{E}(r_{i}^{2})=d_{i}^{2}+\sigma^{2}$
for $1\leq m\leq2$ and $1\leq n\leq2$, the $(m,n)$th element of
the FIM is solely related to $f_{1}\left(d_{i}\right)$ in {\small{}$\text{L}(\bm{\theta})$}
and is expressed as follows
\begin{align}
\left[\mathcal{F}\right]_{m,n}=-\mathbb{E}\left(\frac{\partial^{2}\text{L}(\mathbf{\bm{\theta}})}{\partial\theta_{m}\partial\theta_{n}}\right) & =-\sum_{i=1}^{d\left(\mathbf{p}^{k}\right)}\mathbb{E}\left(\frac{\partial^{2}f_{1}\left(d_{i}\right)}{\partial d_{i}^{2}}\frac{\partial^{2}d_{i}}{\partial p_{m}^{k}\partial p_{n}^{k}}\right)\nonumber \\
=\frac{1}{\sigma^{2}}\sum_{i\,=\,1}^{d\left(\mathbf{p}^{k}\right)}\frac{\partial^{2}d_{i}}{\partial p_{m}^{k}\partial p_{n}^{k}} & =\sum_{i\,=\,1}^{d\left(\mathbf{p}^{k}\right)}\frac{(p_{m}^{k}-s_{im})(p_{n}^{k}-s_{in})}{\sigma^{2}\|\mathbf{p}^{k}-\bm{s}_{i}\|^{2}}.
\end{align}

Additionally, we have, for $1\leq m\leq2<n\leq2\left(\mathbf{n}_{k-1}+1\right)$,
define $n=2n_{1}+n_{2}$, thus $w_{n}$ in estimator $\bm{\theta}$
represents the $n_{2}^{th}$ element of the $n_{1}^{th}$ node of
all nodes from the k-1 level and below, where $n_{1}=1,\ldots,\mathbf{n}_{k-1}$
and $1\leq n_{2}\leq2$. The $(m,n)$th element of the FIM is written
as follows 
\begin{align}
\left[\mathcal{F}\right]_{m,n}=-\mathbb{E}\left(\frac{\partial^{2}\text{L}(\mathbf{\bm{\theta}})}{\partial\theta_{m}\partial\theta_{n}}\right) & =-\mathbb{E}\left(\frac{\partial^{2}}{\partial d_{n_{1}}^{2}}f_{1}\left(d_{n_{1}}\right)\frac{\partial^{2}d_{n_{1}}}{\partial p_{m}^{k}\partial p_{n_{2}}^{k}}\right)\nonumber \\
=\frac{1}{\sigma^{2}}\left(\frac{\partial^{2}d_{n_{1}}}{\partial p_{m}^{k}\partial p_{n_{2}}^{k}}\right) & =\frac{(p_{m}^{k}-s_{n_{1}m})(p_{n_{2}}^{k}-s_{n_{1}n_{2}})}{\sigma^{2}\|\mathbf{p}^{k}-\bm{\mathbf{s}}_{n_{1}}\|^{2}}
\end{align}
where $\mathbf{p}^{k}$ and $\bm{s}_{n_{1}}$ are cross-level nodes.
Similarly , for $3\leq m\leq n\leq2\left(\mathbf{n}_{k-1}+1\right)$,
define $m=2m_{1}+m_{2}$ represents the $m_{2}^{th}$ element of the
$m_{1}^{th}$ node in estimator $\mathbf{W}$. The $(m,n)$th element
of the FIM corresponding to $f_{2}\left(d_{ij}\right)$ in $\text{L}\left(\mathbf{\bm{\theta}}\right)$
is expressed as follows
\begin{align}
\left[\mathcal{F}\right]_{m,n} & =-\mathbb{E}\left(\frac{\partial^{2}}{\partial d_{m_{1}n_{1}}^{2}}f_{2}\left(d_{m_{1}n_{1}}\right)\frac{\partial^{2}d_{m_{1}n_{1}}}{\partial s_{m_{1}m_{2}}\partial s_{m_{1}n_{2}}}\right)\\
=\frac{1}{\sigma^{2}} & \left(\frac{\partial^{2}d_{m_{1}n_{1}}}{\partial s_{m_{1}m_{2}}\partial s_{m_{1}n_{2}}}\right)=\frac{\begin{smallmatrix}\left(s_{m_{1}m_{2}}-s_{n_{1}m_{2}}\right)\left(s_{m_{1}n_{2}}-s_{n_{1}n_{2}}\right)\end{smallmatrix}}{\sigma^{2}\|\bm{\mathbf{s}}_{m_{1}}-\bm{\mathbf{s}}_{n_{1}}\|^{2}}\nonumber 
\end{align}
where $\mathbf{s}_{m_{1}}$ and $\mathbf{s}_{n_{1}}$ are cross-level
nodes. If $1\leq m_{1}=n_{1}\leq\left(\mathbf{n}_{k-1}-\mathbf{n}_{0}\right)$,
we have
\begin{align}
\left[\mathcal{F}\right]_{m,n} & =-\sum_{i=1}^{d\left(\bm{\mathbf{s}}_{m_{1}}\right)}\mathbb{E}\left(\frac{\partial^{2}}{\partial d_{im_{1}}^{2}}f_{2}\left(d_{im_{1}}\right)\frac{\partial^{2}d_{im_{1}}}{\partial s_{im_{2}}\partial s_{m_{1}n_{2}}}\right)\nonumber \\
 & =\sum_{i=1}^{d\left(\bm{\mathbf{s}}_{m_{1}}\right)}\frac{\left(s_{m_{1}m_{2}}-s_{im_{2}}\right)\left(s_{m_{1}n_{2}}-s_{in_{2}}\right)}{\sigma^{2}\|\bm{\mathbf{s}}_{m_{1}}-\bm{\mathbf{s}}_{i}\|^{2}}
\end{align}
where $\mathbf{s}_{m_{1}}$ is a node between level 1 and level k-1.
Moreover, if $\left(\mathbf{n}_{k-1}-\mathbf{n}_{0}\right)\leq m_{1}=n_{1}\leq\mathbf{n}_{k-1}$
and $m_{2}=n_{2}$ then point $\mathbf{s}_{m_{1}}^{0}$ is a base
anchor, associated with both $f_{2}\left(d_{ij}\right)$ and $f_{3}\left(\mathbf{s}_{i}^{0}\right)$.
The $(m,n)$th element of the FIM can be represented as follows
\begin{align}
\left[\mathcal{F}\right]_{m,n} & =-\sum_{i=1}^{d\left(\bm{\mathbf{s}}_{m_{1}}\right)}\mathbb{E}\left(\frac{\partial^{2}f_{2}\left(d_{im_{1}}\right)}{\partial s_{im_{2}}\partial s_{m_{1}n_{2}}}+\frac{\partial^{2}f_{3}\left(\mathbf{s}_{m_{1}}^{0}\right)}{\partial^{2}\mathbf{s}_{m_{1}m_{2}}}\right)\nonumber \\
 & =\sum_{i=1}^{d\left(\bm{\mathbf{s}}_{m_{1}}\right)}\frac{(s_{m_{1}m_{2}}-s_{im_{2}})^{2}}{\sigma^{2}\|\bm{\mathbf{s}}_{m_{1}}-\bm{\mathbf{s}}_{i}\|^{2}}+\frac{\partial^{2}\|\mathbf{s}_{m_{1}}^{0}-\tilde{\mathbf{s}}_{m_{1}}^{0}\|^{2}}{2\delta^{2}\partial^{2}\mathbf{s}_{m_{1}m_{2}}}\nonumber \\
 & =\sum_{i=1}^{d\left(\bm{\mathbf{s}}_{m_{1}}\right)}\frac{\left(s_{m_{1}m_{2}}-s_{im_{2}}\right)^{2}}{\sigma^{2}\|\bm{\mathbf{s}}_{m_{1}}-\bm{\mathbf{s}}_{i}\|^{2}}+\frac{1}{\delta^{2}}.
\end{align}

\section{PROOF OF REMARK \ref{rem:update CRLB}\label{sec:PROOF-OF-REMARK}}

First, consider a target node $\mathbf{p}$ alongside its dynamic
anchors $\bm{\mathbf{s}}_{i}$, where $i=1,2,...,n$. For the estimator
$\bm{\theta}=\left(\mathbf{p},\bm{\mathbf{s}}_{1},\bm{\mathbf{s}}_{2},...,\bm{\mathbf{s}}_{n}\right)$,
we compare the FIM in two separate estimations. When $\bm{\mathbf{s}}_{i}$
serves as the preceding-level target point, its covariance matrix
is represented as $\boldsymbol{\Sigma}_{si}$, while the corresponding
CRLB is indicated by $\sigma_{ci}^{2}$. Initialize the covariance
matrix of the target point as $\boldsymbol{\Sigma}_{p}$. Therefore,
the FIM for the estimator $\bm{\theta}$ before localizing target
$\mathbf{p}$ is as follows
\begin{equation}
\mathcal{F}_{\text{in}}=\text{diag}\left[\boldsymbol{\Sigma}_{p}^{-1},\boldsymbol{\Sigma}_{s1}^{-1},\boldsymbol{\Sigma}_{s2}^{-1},...,\boldsymbol{\Sigma}_{sn}^{-1}\right].
\end{equation}

When $\bm{\mathbf{s}}_{i}$ serves as the dynamic anchor for target
$\mathbf{p}$, the updated FIM is as follows
\begin{equation}
\mathcal{F}_{\text{up}}=-\mathbb{E}\left(\frac{\partial^{2}\text{L}(\mathbf{\bm{\theta}})}{\partial\theta_{m}\partial\theta_{n}}\right)=\mathcal{F}_{\text{in}}+\bm{G}_{d}
\end{equation}
where $\bm{G}_{d}$ is the TOA-related sub-matrices. Based on the
Section \ref{sec:Cram=0000E9r=002013Rao-Lower-Bound}, we have
\begin{equation}
\bm{G}_{d}=\left[\Delta\boldsymbol{p},\Delta\boldsymbol{s}_{1},...,\Delta\boldsymbol{s}_{n}\right]^{\text{T}}\left[\Delta\boldsymbol{p},\Delta\boldsymbol{s}_{1},...,\Delta\boldsymbol{s}_{n}\right]
\end{equation}
where
\begin{align}
\Delta\boldsymbol{p} & =\left[\sum_{i\,=\,1}^{n}\frac{\mathbf{p}_{1}-\bm{\mathbf{s}}_{i1}}{\sigma^{2}\|\mathbf{p}-\bm{\mathbf{s}}_{i}\|^{2}},\sum_{i\,=\,1}^{n}\frac{\mathbf{p}_{2}-\bm{\mathbf{s}}_{i2}}{\sigma^{2}\|\mathbf{p}-\bm{\mathbf{s}}_{i}\|^{2}}\right],\nonumber \\
\Delta\boldsymbol{s}_{i} & =\left[\frac{\mathbf{p}_{1}-\bm{\mathbf{s}}_{i1}}{\sigma^{2}\|\mathbf{p}-\bm{\mathbf{s}}_{i}\|^{2}},\frac{\mathbf{p}_{2}-\bm{\mathbf{s}}_{i2}}{\sigma^{2}\|\mathbf{p}-\bm{\mathbf{s}}_{i}\|^{2}}\right],\,i=1,...,n.
\end{align}

Note that the matrix $\bm{G}_{d}$ is positive definite. Consequently,
we have
\begin{equation}
\mathcal{F}_{\text{up}}\succ\mathcal{F}_{\text{in}}
\end{equation}

The CRLB is the diagonal elements in the inverse FIM. Therefore, the
relationship between the CRLBs of the updated estimation and the initial
one can be 
\begin{equation}
(\sigma_{ci}^{2})^{\prime}<\sigma_{ci}^{2}
\end{equation}

Thus, we have proven that the accuracy of the updated dynamic anchors
is higher than the previous ones.

\bibliographystyle{IEEEtran}
\bibliography{transyxk.bbl}

\begin{IEEEbiography}[{\fbox{\begin{minipage}[t][1.25in]{1in}%
Replace this box by an image with a width of 1\,in and a height of
1.25\,in!%
\end{minipage}}}]{Your Name}
 All about you and the what your interests are.
\end{IEEEbiography}

\begin{IEEEbiographynophoto}{Coauthor}
Same again for the co-author, but without photo
\end{IEEEbiographynophoto}

\end{document}